\documentclass[useAMS,
usenatbib,
onecolumn
]{mn2e}
\usepackage{amssymb, amsmath}

\usepackage{epsf}
\usepackage{bm}

\title[Dissipation in relativistic superfluid neutron stars]
{Dissipation in relativistic superfluid neutron stars}
\author[M. E. Gusakov et al.]
{M.~E.~Gusakov$^1$\thanks{gusakov@astro.ioffe.ru},
 E.~M.~Kantor$^{1,2}$\thanks{kantor@mail.ioffe.ru},
A.~I.~Chugunov$^1$\thanks{andr.astro@mail.ioffe.ru},
L.~Gualtieri$^3$\thanks{leonardo.gualtieri@roma1.infn.it}\\
$^1$ Ioffe Physical-Technical Institute of the Russian Academy of
Sciences, Polytekhnicheskaya 26, 194021 Saint-Petersburg, Russia\\
$^2$ Saint-Petersburg State Polytechnical University,
Polytekhnicheskaya 29, 195251 St.-Petersburg, Russia\\
$^3$ Dipartimento di Fisica, ⌠Sapienza■ Universit$\acute{{\rm a}}$ 
di Roma \& Sezione INFN Roma1, Piazzale Aldo Moro 5, 00185, Roma, Italy}

\begin{document}

\date{Accepted 2012 xxxx. Received 2012 xxxx;
in original form 2012 xxxx}

\pagerange{\pageref{firstpage}--\pageref{lastpage}} \pubyear{2012}

\maketitle

\label{firstpage}

%
\begin{abstract}

We analyze damping of oscillations 
of general relativistic superfluid neutron stars.
To this aim we extend the method of decoupling 
of superfluid and normal oscillation modes
first suggested in [Gusakov \& Kantor PRD 83, 081304(R) (2011)]. 
All calculations are made self-consistently
within the finite temperature superfluid hydrodynamics.
The general analytic formulas are derived for damping times due to
the shear and bulk viscosities. 
These formulas describe both normal
and superfluid neutron stars and are valid for oscillation modes of
arbitrary multipolarity.
We show that:
($i$) use of the ordinary one-fluid hydrodynamics
is a good approximation, 
for most of the stellar temperatures, 
if one is interested in 
calculation of the damping times of normal $f$-modes;
($ii$) for radial and $p$-modes such an approximation is poor;  
($iii$) the temperature dependence of damping times
undergoes a set of rapid changes associated 
with resonance coupling of neighboring oscillation modes. 
The latter effect can substantially accelerate viscous damping of
normal modes in certain stages 
of neutron-star thermal evolution.
\end{abstract}
%

\begin{keywords}
stars: neutron -- stars: oscillations -- stars: interiors.
\end{keywords}

\maketitle

\section{Introduction}
\label{Sec:Intro}

Neutron stars (NS) are compact objects 
with the mass $M \sim M_{\odot}$,
circumferential radius $R \sim 10$~km, 
and the central density $\rho_{\rm c}$ 
several times higher than the nuclear density
$\rho_0 \approx 2.8 \times 10^{14}$~g cm$^{-3}$.
They are interesting because of extreme conditions in 
their interiors and a wide variety 
of associated astrophysical phenomena. 
In particular, internal instabilities 
or external perturbations can excite NS oscillations, 
which are potentially detectable by the next-generation 
gravitational wave interferometers
(see, e.g., \citealt*{ak01,andersson03, owen10}).
It is very probable,
that quasiperiodic oscillations 
of electromagnetic radiation observed 
in the tails of the giant gamma-ray flares 
are connected with oscillations in NS crust
(e.g., \citealt{israel_etal05, sw05, sw06, ws07}), 
and that seismology would become a significant source of information
about NSs in the nearest future (\citealt{abbott_etal07, watts11, afjkkrrz11}).

For the correct interpretation of already existing and 
future observations one requires a well-developed 
theory of oscillating NSs.
It should, in particular:
($i$) be based on the general relativity theory, 
since NSs are relativistic objects;
($ii$) employ an adequate model of superdense matter, 
including realistic equation of state 
and parameters of baryon superfluidity;
($iii$) correctly account for the effects
of baryon superfluidity 
on the hydrodynamics of NS matter.

Let us discuss briefly a (key) role 
of superfluidity.
According to numerous microscopic calculations (see, e.g., \citealt{ls01}), 
baryon matter in the internal layers of neutron stars 
becomes superfluid at $T \la 10^8$--$10^{10}$~K. 
It is very difficult to interpret
the observational data on pulsar glitches 
(see, e.g., \citealt{ch08}) and cooling of NSs (\citealt*{yls99, yp04}) 
without invoking baryon superfluidity.
Recent real-time observations of cooling NS 
in Cassiopea A supernova remnant (\citealt{hh10})
also present a strong argument in favor of 
the existence of baryon superfluidity 
in the NS core.
The observations were explained by \cite{syhhp11} and \cite{ppls11} 
within a scenario, 
suggested for the first time
in \cite{gkyg04} and \cite{plps04},
and assuming mild neutron superfluidity
(with maximum neutron critical temperatures
$T_{\rm cn \, max} \sim 7 \div 9 \times 10^8$~K)
and strong proton superconductivity
(with maximum proton critical temperatures
$T_{\rm cp \, max} \ga 2\div3 \times 10^9$~K) in the NS core.

Combined analysis of all the three factors ($i$)--($iii$) 
is a formidable task for the oscillation theory so 
in the literature they were considered successively.
The foundations of the relativistic theory 
of stellar oscillations
were laid fifty years ago by 
\cite{chandrasekhar64} and \cite{tc67} 
and were further developed in many subsequent papers
(see, e.g., 
\citealt*{it73, di73, ld83, dl85, cl87, cls90, cf91, ks92, yl03b, lac08}
and a review of \citealt{ks99}). 
When studying oscillations of NSs, 
most of these works considered ordinary 
one-fluid relativistic hydrodynamics.

Meanwhile it is well known that superfluidity 
leads to appearance of additional velocity fields,
describing the superfluid degrees of freedom 
(e.g., \citealt{kl82, khalatnikov89, ck92}).
This substantially complicates the 
hydrodynamics of NS matter, making it multi-fluid
(\citealt{mendell91a, mendell91b, ga06}). 
In addition, superfluidity affects
the kinetic coefficients 
(such as bulk and shear viscosities) 
and also requires additional viscous coefficients 
to be introduced (see \citealt{gusakov07, gk08} for details).

Oscillations of superfluid NSs 
have been studied actively 
only in the last two decades
(see, e.g., \citealt*{lee95, lm00, pr02, yl03c, pca04, sa09, wll09, pa11, pa12}), 
starting from the pioneering papers by \cite{epstein88} and \cite{lm94}.
However, most of these works neglect general relativity effects 
and employ zero temperature ($T=0$) limit of superfluid hydrodynamics 
(i.e., hydrodynamics, applicable only at $T=0$).
Within the general relativity theory 
oscillations 
were discussed 
only by
\cite*{cll99, acl02, yl03b}; \cite{ga06, lac08, kg11, cg11}, 
but most of these works used zero temperature hydrodynamics.
Moreover, in some of these papers (e.g., \citealt{acl02, lac08}) 
the presence of superfluid
component was modeled by an artificial (polytropic)
equation of state which does not represent any specific 
microphysical model. 

Only in the recent papers \cite{ga06,kg11,cg11} 
an attempt was made 
to self-consistently calculate the oscillation spectra
using a realistic model of superdense matter and 
allowing for the effects of finite stellar temperatures.
It was shown that in many cases 
an approximation $T=0$ is not justified 
and, moreover, 
it can lead to {\it qualitatively incorrect} 
results (\citealt{kg11,cg11}).

Of particular interest is the question of
how superfluidity influences dissipation
of neutron star oscillations.  
It is of extreme importance, for instance, 
for understanding physical conditions under which 
a rotating NS becomes unstable 
with respect to
excitation of various oscillations 
(e.g., $r$-modes), 
and for estimating gravitational radiation from such stars 
(e.g., \citealt{ak01}).

There were several serious and successful attempts 
to allow for the effects of superfluidity when studying the 
dissipation of oscillations in NSs
(see, e.g., \citealt*{lm95, lm00, ly03a, hap09, agh09, ha10, pg12}),
but {\it all of them} considered Newtonian stars and 
used the $T=0$ superfluid hydrodynamics. 
The self-consistent analysis of dissipation 
in superfluid NSs was only recently performed
for a simple case of a radially oscillating NS (\citealt{kg11}).

The aim of the present paper 
is to fill this gap 
and to consider, 
for the first time,
dissipation of nonradial oscillations 
in general relativistic superfluid NSs
employing realistic microphysics input 
with accurate treatment of the effects 
of finite stellar temperatures.

The paper is organized as follows.
Relativistic superfluid 
hydrodynamics is briefly reviewed 
in Sec.\  \ref{Sec:hydro}.
Sec.\ \ref{Sec:basic} 
discusses an unperturbed star
and introduces variables 
describing small deviations
of NS from equilibrium.
In Sec.\ \ref{Sec:damping_general} 
we derive expressions 
for the oscillation energy
and its dissipation rates 
due to bulk and shear viscosities.
In Sec.\ \ref{SubSec:osc}
the equations 
that govern oscillations 
of superfluid NSs
are explicitly written out. 
Sec.\ \ref{Sec:approach} describes 
the approach to study dissipation 
of superfluid NS oscillations.
This approach is applied
for a detailed numerical analysis of 
realistic models 
of oscillating neutron stars in Sec.\ \ref{Sec:Results}.
Sec.\ \ref{Sec:summary} presents a summary of our results.

In what follows, we use the system of units 
in which $c=k_\mathrm{B}=1$, 
where $c$ is the speed of light 
and $k_\mathrm{B}$ is the Boltzmann constant.

\section{Dissipative superfluid hydrodynamics}
\label{Sec:hydro}
In this paper we consider, for simplicity, 
npe-matter in NS cores, 
that is matter composed of 
neutrons (n), protons (p), and electrons (e). 
Because both protons and neutrons 
can be in the superfluid state, 
one has to use the relativistic hydrodynamics 
of superfluid mixtures 
to study oscillations of NSs. 
Here we briefly discuss 
the corresponding equations
to establish notations and to make the presentation 
more self-contained. 
Our consideration
closely follows the papers by
\cite{ga06, gusakov07} and, especially, \cite{kg11}.
The reader is referred to these works for more details.

The main distinctive feature of superfluid hydrodynamics 
is the presence of several velocity fields in the mixture. 
In our case, these are the four-velocity $u^{\mu}$ 
of the `normal' (nonsuperfluid) component of matter 
(electrons and Bogoliubov excitations of neutrons and protons) 
as well as the `four-velocities' of superfluid neutrons $v_{s({\rm n})}^{\mu}$ 
and superfluid protons $v_{s({\rm p})}^{\mu}$. 
In what follows instead of the velocities
$v_{s({\rm n})}^{\mu}$ and $v_{s({\rm p})}^{\mu}$
it will be convenient to use the four-vectors 
$w^{\mu}_{(i)} = \mu_i[v^{\mu}_{s(i)}-u^{\mu}]$, 
where $\mu_i$ is the relativistic chemical potential 
for particle species $i={\rm n}$ or ${\rm p}$.
A presence of several velocity fields modifies the expressions 
for the current densities of neutrons $j^{\mu}_{({\rm n})}$ 
and protons $j^{\mu}_{({\rm p})}$,
\begin{equation}
j^{\mu}_{(i)} = n_i u^{\mu} + Y_{ik} w^{\mu}_{(k)}
\label{jnp}
\end{equation}
in comparison with  
the standard expression $j^{\mu}_{(i)} = n_i u^{\mu}$.
The electron current density $j^{\mu}_{({\rm e})}$ 
has a standard form,
\begin{equation}
j^{\mu}_{({ {\rm e}})} = n_{\rm e} u^{\mu}.
\label{je}
\end{equation}
Here and below the subscripts $i$ and $k$ 
refer to nucleons: $i$, $k = {\rm n}$, ${\rm p}$; 
$n_l$ is the number density of particle species 
$l={\rm n}$, ${\rm p}$, ${\rm e}$.
Unless otherwise stated the summation 
is assumed over the repeated nucleon indices 
$i$, $k$ and over the spacetime indices 
$\mu$, $\nu$, $\ldots$ (Greek letters).
In Eq.\ (\ref{jnp})
$Y_{ik}$ 
is the relativistic entrainment matrix, 
which is a generalization of the concept 
of superfluid density (see, e.g., \citealt{khalatnikov89}) 
to the case of relativistic mixtures. 
In the nonrelativistic theory, 
a similar matrix was first considered 
by \cite{ab75}. 
The matrix $Y_{ik}$
is symmetric, $Y_{ik}=Y_{ki}$,
and is expressed in terms 
of the Landau parameters $F_1^{ik}$ 
of asymmetric nuclear matter 
and universal functions of temperature, 
$\Phi_i$, as described in \cite*{gkh09b}. 
In beta-equilibrium
it can be presented as 
a function of density $\rho$
and the combinations $T/T_{c{\rm n}}$ and $T/T_{c{\rm p}}$:
$Y_{ik}=Y_{ik}(\rho, T/T_{c{\rm n}}, T/T_{c{\rm p}})$,
where $T$ is the temperature;
$T_{c{\rm n}}(\rho)$ and $T_{c{\rm p}}(\rho)$ 
are the density-dependent 
neutron and proton critical temperatures, respectively.
If, for example, $T > T_{c {\rm n}}$ then all neutrons are normal.
The important property of the matrix $Y_{ik}$ is that
for any nonsuperfluid species $l={\rm n}$ or ${\rm p}$, 
the corresponding elements $Y_{l k}$ of this matrix vanish.

In the present paper we consider NS oscillations, 
whose frequencies are well below the electron and proton
plasma frequencies.
In that case the quasineutrality condition,
$n_{\rm e}=n_{\rm p}$, 
should 
hold
in an oscillating star, 
from which it follows 
(for a nonrotating non-magnetized NS)
$j^{\mu}_{({\rm p})}=j^{\mu}_{({\rm e})}$ or,
in view of (\ref{jnp}) and (\ref{je}),
\begin{equation}
Y_{pk} w^{\mu}_{(k)}=0.
\label{quasineutrality}
\end{equation}
Below we assume that this condition is always satisfied.
It relates the four-vectors 
$w^{\mu}_{({\rm n})}$ and $w^{\mu}_{({\rm p})}$.

In what follows, along with $u^{\mu}$ and $w^{\mu}_{(i)}$ 
it will be convenient to introduce the quantity $X^{\mu}$, 
describing superfluid degrees of freedom, 
as well as the quantity 
which we call the `baryon four-velocity' $U^\mu_{({\rm b})}$
(notice, however, 
that it is not a four-velocity in the usual sense, 
because generally $U^{\mu}_{({\rm b})} U_{({\rm b}) \, \mu} \neq -1$, 
see Eq.\ (\ref{normub}) and the footnote $^{\ref{umuumu}}$ below).
They are defined by the formulas
\begin{eqnarray}
X^{\mu} &=& \frac{Y_{{\rm n}k} w^{\mu}_{(k)}}{n_{\rm b}},
\label{X0}\\
U^\mu_{({\rm b})} &=& u^{\mu}+X^{\mu},
\label{ub}
\end{eqnarray}
where $n_{\rm b}=n_{\rm n}+n_{\rm p}$ 
is the baryon number density.
Notice that, as follows from 
Eqs.\ (\ref{jnp})--(\ref{quasineutrality}),
the baryon current density
$j^{\mu}_{({\rm b})}=j^{\mu}_{({\rm n})}+j^{\mu}_{({\rm p})}$ 
is related to $U^\mu_{({\rm b})}$ by the standard equation, 
\begin{equation}
j^{\mu}_{({\rm b})}=n_{\rm b} \,
U^{\mu}_{({\rm b})},
\label{jb}
\end{equation}
while $j_{\rm (e)}^{\mu}$ equals
\begin{equation}
j^{\mu}_{\rm (e)}=n_{\rm e}  \left[U_{({\rm b})}^{\mu} - X^{\mu} \right].
\label{je2}
\end{equation}

Together with the quasineutrality condition 
($n_{\rm e}=n_{\rm p}$) and Eq. (\ref{quasineutrality}), 
the equations of superfluid hydrodynamics include (\citealt{gusakov07}):

($i$) Continuity equations for baryons (b) and electrons (e),
\begin{eqnarray}
j^{\mu}_{ ({\rm b}) ; \, \mu} &=& 0, 
\label{cont_b}\\
j^{\mu}_{({\rm e}) ; \, \mu} &=& 0;
\label{cont_e}
\end{eqnarray}

($ii$) Energy-momentum conservation
\begin{eqnarray}
T^{\mu \nu}_{; \,\mu} &=& 0, 
\label{Tmunu_cons}\\
T^{\mu \nu} &=& (P+\varepsilon) \, u^{\mu} u^{\nu} + P g^{\mu \nu} 
+ Y_{ik} \left( w^{\mu}_{(i)} w^{\nu}_{(k)} + \mu_i \, w^{\mu}_{(k)} u^{\nu} 
+ \mu_k \, w^{\nu}_{(i)} u^{\mu} \right)+\tau^{\mu\nu},
\label{Tmunu}\\
\tau^{\mu\nu} &=&  
- \eta \, H^{\mu \gamma} \, H^{\nu \delta} \,\, 
\left( u_{\gamma ; \delta} + u_{\delta; \gamma}  
- {2 \over 3} \,\, g_{\gamma \delta} \,\, 
u^{\varepsilon}_{;\varepsilon}  \right)
- \xi_{1 {\rm n}} \, H^{\mu \nu} \,
\left[ Y_{{\rm n} k} w^{\gamma}_{(k)} \right]_{; \gamma}
- \xi_2 \, H^{\mu \nu} \, u^{\gamma}_{;\gamma};
\label{taumunu}
\end{eqnarray}

($iii$) Potentiality condition 
for superfluid motion of neutrons
\begin{eqnarray}
\partial_{\nu} \left[ w_{({\rm n}) \mu} 
+ (\mu_{\rm n} + \varkappa_{\rm n}) u_{\mu} \right]
&=& \partial_{\mu} \left[ w_{({\rm n}) \nu} 
+(\mu_{\rm n} + \varkappa_{\rm n}) u_{\nu} \right],
\label{wmu22}\\
\varkappa_{\rm n} &=&  
- \xi_{3{{\rm n}}} \, \left[ Y_{{\rm n}k} w^{\mu}_{(k)} \right]_{;\mu}
- \xi_{4{\rm n}} \, u^{\mu}_{;\mu};
\label{kappa_n}
\end{eqnarray}
as well as ($iv$) the second law of thermodynamics
\begin{equation}
{\rm d} \varepsilon =  T \, {\rm d} S + \mu_e \, {\rm d} n_e + \mu_i \, {\rm d} n_i 
+ \frac{Y_{ik}}{2} \, {\rm d} \left( w^{\alpha}_{(i)} w_{(k) \alpha} \right).
\label{2ndlaw}
\end{equation}
In formulas (\ref{cont_b})--(\ref{2ndlaw})
$g^{\mu \nu}$ is the metric tensor;
$H^{\mu \nu} \equiv g^{\mu\nu}+u^{\mu}u^{\nu}$; 
$\partial_\mu \equiv \partial/(\partial x^\mu)$;
$P$, $\varepsilon$, $S$, and $\mu_{\rm e}$ 
are the pressure, energy density, entropy density, 
and relativistic electron chemical potential, 
respectively.
These quantities are related by the formula 
\begin{equation}
P=-\varepsilon+ \mu_{\rm e} n_{\rm e} + \mu_i n_i + TS.
\label{PPP}
\end{equation}
Finally, $\eta$ is the shear viscosity coefficient and
$\xi_{1{\rm n}}$,  $\xi_{2}$, $\xi_{3{\rm n}}$, $\xi_{4{\rm n}}$ 
are the bulk viscosity coefficients.
Because of the Onsager symmetry principle, one has
\begin{equation}
\xi_{1{\rm n}}=\xi_{4{\rm n}}.
\label{onsager}
\end{equation}
Moreover, if the bulk viscosities are generated solely by the
direct or modified URCA processes, 
one has an additional constraint (\citealt{gusakov07})
\begin{equation}
\xi_{1{\rm n}}^2=\xi_2 \xi_{3{\rm n}}.
\label{xi_cond}
\end{equation}
In the absence of superfluidity the only nonzero
coefficient is $\xi_2$ -- the ordinary bulk viscosity.

To close the system describing superfluid hydrodynamics 
one should put two additional constraints 
on the four-vectors $u^{\mu}$ and $w^{\mu}_{({\rm n})}$,
\begin{eqnarray}
u_{\mu}u^{\mu}&=&-1,
\label{normalization}\\
u_{\mu} w^{\mu}_{({\rm n})}&=&0.
\label{uw}
\end{eqnarray}
The first constraint is the standard normalization condition
while the second one indicates that the comoving frame,
in which we measure various thermodynamic quantities 
(e.g., $n_i$, $\varepsilon$, $\ldots$), 
is defined by the condition $u^{\mu}=(1,0,0,0)$ (\citealt{ga06, gusakov07}).
Using Eqs.\ (\ref{jnp}), (\ref{Tmunu}), (\ref{taumunu}), (\ref{normalization}), and (\ref{uw}) 
one then immediately finds that $n_{l}=-u_{\mu} j^{\mu}_{(l)}$ ($l=$n, p, e)
and $\varepsilon = u_{\mu} u_{\nu} T^{\mu \nu}$.

Making use of the hydrodynamics described above,
one can derive the entropy generation equation, 
valid for superfluid matter.
Following the derivation of the similar equation (33) in \cite{gusakov07},
one arrives at
\begin{equation}
S^{\mu}_{;\mu} = 
- {\varkappa_{\rm n}  \over T} \,\, \left[ Y_{{\rm n}k} w^{\mu}_{(k)} \right]_{;\mu}
-\tau^{\mu \nu} \,\, \left( {u_{\nu} \over T} \right)_{;\mu}
\label{entropy2}
\end{equation}
where the entropy density current $S^{\mu}$ is 
\footnote{
Notice that, in \cite{gusakov07} there is an additional
term in the expression for $S^{\mu}$, so that 
\begin{equation}
S^{\mu} = S u^{\mu} 
- {u_{\nu} \over T} \,\, \tau^{\mu \nu}
- {\varkappa_{\rm n} \over T} \,\, Y_{{\rm n}k} w^{\mu}_{(k)}.
\nonumber\\
\end{equation}
The last term here appears naturally 
in the entropy generation equation.
However, strictly speaking, 
it is small and should be neglected
if one takes into account only the {\it largest} dissipative terms 
in the equations of superfluid hydrodynamics 
(this is the standard approximation; 
see \citealt{gusakov07} and $\S$140 of \citealt{ll87} 
for an explanation of what we mean by the `largest terms').
It remains to note that the terms similar 
to the last term in the expression for $S^{\mu}$
also appear in the most general form of 
the nonrelativistic superfluid dissipative hydrodynamics 
formulated by Clark 
(for details see the book 
by \citealt{putterman74}).
}
%
\begin{equation}
S^{\mu} = S u^{\mu} 
- {u_{\nu} \over T} \,\, \tau^{\mu \nu}. 
\label{S_mu2}
\end{equation}
When writing (\ref{entropy2}) 
we neglected small dissipative terms, 
as it is discussed in \cite{gusakov07}.
Introducing
\begin{eqnarray}
Q_{\rm bulk} &\equiv& \left\{
\sqrt{\xi_{3{\rm n}}} \,\,\left[Y_{{\rm n}k} w^{\mu}_{(k)}\right]_{;\mu}
+\sqrt{\xi_2} \,\, u^{\mu}_{;\mu} \right\}^2,
\label{bulk}\\
Q_{\rm shear} &\equiv& \eta \, H^{\mu \gamma} \, H^{\nu \delta} \,\, 
\left( u_{\gamma ; \delta} + u_{\delta; \gamma}  
- {2 \over 3} \,\, g_{\gamma \delta} \,\, 
u^{\varepsilon}_{;\varepsilon}  \right) u_{\nu ;\mu}
\nonumber\\
&=& \frac{\eta}{2} \, H^{\mu \gamma} \, H^{\nu \delta} \,\, 
\left( u_{\gamma ; \delta} + u_{\delta; \gamma}  
- {2 \over 3} \,\, g_{\gamma \delta} \,\, 
u^{\varepsilon}_{;\varepsilon}  \right)
\left( u_{\nu ; \mu} + u_{\mu; \nu}  
- {2 \over 3} \,\, g_{\mu \nu} \,\, 
u^{\varepsilon}_{;\varepsilon}  \right),
\label{shear}
\end{eqnarray}
Eq. (\ref{entropy2}) can be rewritten as
\begin{equation}
T (S u^{\mu})_{;\mu}=Q_{\rm bulk}+Q_{\rm shear}.
\label{entropy3}
\end{equation}
To derive Eq.\ (\ref{entropy3}) 
we used Eqs.\ (\ref{onsager}) and (\ref{xi_cond}), 
as well as the fact that for the tensor (\ref{taumunu}) 
$\tau^{\mu\nu} \, u_{\nu}=0$
\footnote{This equality holds true only if one neglects
the thermal conductivity, 
as we assume in Eq.\ (\ref{taumunu}).}.

\section{Basic equations}
\label{Sec:basic}

\subsection{An unperturbed star}
\label{SubSec:unperturbed}

An equilibrium configuration 
of a nonrotating superfluid NS 
was analyzed in detail in section 3 of \cite{ga06}.
Here we present only the main results of this analysis, 
which will be used in what follows.

The metric of a spherically symmetric, 
nonrotating NS in equilibrium has the form
\begin{equation}
-{\rm d} s^2 \equiv g_{\alpha\beta}^{(0)} {\rm d} x^{\alpha} {\rm d} x^{\beta} =
 - {\rm e}^{\nu} {\rm d} t^2+{\rm e}^{\lambda}{\rm d} r^2
+ r^2 ({\rm d} \theta^2+{\rm sin^2 \theta} \, {\rm d}\varphi^2),
\label{ds}
\end{equation}
where $r$, $\theta$, and $\varphi$, 
are the spatial coordinates in the spherical
frame with the origin at the stellar centre; 
$t$ is the time coordinate;
$\nu(r)$ and $\lambda(r)$ are the metric coefficients 
for an unperturbed star.

The four-velocity $u^{\mu}$, 
generally defined as
\begin{equation}
u^{\mu}=\frac{{\rm d}x^{\mu}}{{\rm d}s},
\label{umu}
\end{equation}
in equilibrium equals
\begin{equation}
u^{0}={\rm e}^{-\nu/2}, \quad u^1=u^2=u^3=0.
\label{umueq}
\end{equation}
We assume that in the unperturbed star 
superfluid components are at rest with respect to the normal component.
In that case the four-vectors $w^{\mu}_{(i)}$ satisfy
\begin{equation}
w^{\mu}_{(\rm n)}=w^{\mu}_{(\rm p)}=0.
\label{wmu}
\end{equation}
Using Eqs.\ (\ref{X0}), (\ref{ub}), (\ref{umueq}), and (\ref{wmu}),
one has for the baryon four-velocity
\begin{equation}
U^{0}_{({\rm b})}={\rm e}^{-\nu/2}, 
\quad U^1_{({\rm b})}=U^2_{({\rm b})}=U^3_{({\rm b})}=0.
\label{ubeq}
\end{equation}
In addition, the following conditions of hydrostatic equilibrium 
must hold for an unperturbed star,
\begin{eqnarray}
&&\frac{{\rm d}P}{{\rm d}r}=-\frac{1}{2} \, (P+\varepsilon) \, \frac{{\rm d} \nu}{{\rm d}r},
\label{dpdr}\\
&&\frac{{\rm d}}{{\rm d}r}\left( \mu_{\rm n} {\rm e}^{\nu/2} \right)=0.
\label{mun}
\end{eqnarray}
The last condition should be only used in the stellar region 
where neutrons are superfluid (hereafter the SFL-region).
One can show (\citealt{ga06}), 
that if an unperturbed NS is additionally in beta-equilibrium, 
that is, the imbalance $\delta \mu$ of chemical potentials vanishes,
\begin{equation}
\delta \mu \equiv \mu_{\rm n}-\mu_{\rm p}-\mu_{\rm e}=0,
\label{dmu}
\end{equation}
then the SFL-region must also be in thermal equilibrium, 
with the redshifted internal stellar temperature $T^\infty$
constant over this region,
\begin{equation}
T^{\infty}\equiv T {\rm e}^{\nu/2}={\rm constant}.
\label{Tconst}
\end{equation}
In what follows we assume that the conditions (\ref{dmu}) and (\ref{Tconst})
are satisfied in the entire core of the unperturbed NS.
In the latter case Eq.\ (\ref{PPP}) 
for the equilibrium pressure can be rewritten as
\begin{equation}
P=-\varepsilon+\mu_{\rm n} n_{\rm b} + TS.
\label{PPP2}
\end{equation}
It should also be stressed that, 
as long as we neglected the temperature effects 
when calculating the equilibrium stellar model, 
the hydrostatic structure of the unperturbed superfluid NS
is indistinguishable from that of the normal (nonsuperfluid) 
star of the same mass.

\subsection{Small departures from equilibrium}
\label{SubSec:small}
The metric of a perturbed star can be presented in the form
\begin{equation}
-{\rm d} s^2 \equiv g_{\alpha\beta} {\rm d}x^{\alpha} {\rm d}x^{\beta} =
 (g_{\alpha\beta}^{(0)}+\delta g_{\alpha\beta}) {\rm d}x^{\alpha} {\rm d}x^{\beta}.
\label{ds1}
\end{equation}
From here on the symbol $\delta$ denotes Eulerian perturbations,
so that $\delta g_{\alpha\beta}$ 
corresponds to small metric perturbations 
in the course of stellar oscillations.

Since we study oscillations of a nonrotating nonmagnetized NS 
and neglect the effects of crystalline crust, 
all the perturbations in the system are of even parity 
\footnote{A more detailed argument can be found 
in \cite{tc67}; see also \cite{rw57}.}.
In that case, in the appropriately chosen gauge
$\delta g_{\alpha\beta} \, {\rm d}x^{\alpha} {\rm d}x^{\beta}$
can be written as 
(we follow the notations of \citealt{cls90})
\begin{eqnarray}
\delta g_{\alpha \beta} \, {\rm d}x^{\alpha} {\rm d}x^\beta &=&
-{\rm e}^{\nu} \, r^l \, H_0(r)\, Y_{l}^{m} \, {\rm e}^{{\rm i} \omega t} \, {\rm d}t^2
-2\, {\rm i} \,\omega \,r^{l+1} \,  H_1(r) \, Y_{l}^m \,{\rm e}^{{\rm i} \omega t} \, {\rm d}t {\rm d}r
\nonumber\\
&& -{\rm e}^{\lambda} \, r^l \,  H_2(r) \, Y_{l}^m \,{\rm e}^{{\rm i} \omega t} \, {\rm d}r^2
-r^{l+2} \, K(r) \, Y_{l}^m \,{\rm e}^{{\rm i} \omega t} \,
({\rm d} \theta^2+{\rm sin}^2 \theta\, {\rm d} \varphi^2).
\label{metric2}
\end{eqnarray}
In Eq.\ (\ref{metric2}) we assumed 
that all the perturbations depend on $t$ as ${\rm e}^{{\rm i} \omega t}$.
In addition, we already expanded the perturbations 
into series in spherical harmonics $Y_{l}^m$, 
and consider a single harmonic with fixed $l$ and $m$.
The unknown functions $H_0$, $H_1$, $H_2$, and $K$ 
depend on $r$ only, and 
should be determined from the linearized Einstein equations, 
describing NS oscillations (see Sec.\ \ref{SubSec:osc}).
Depending on $l$ the gauge of the metric 
can be further specialized (e.g., \citealt{cls90}).
Namely, one can choose the gauge such that 
for $l=0$ (radial oscillations) $H_1=K=0$;
for $l=1$ (dipole oscillations) $K=0$; 
for $l\geq 2$ $H_0=H_2$.

As follows from the definition (\ref{umu}),
in the perturbed star the four-velocity $u^{\mu}$ 
of the normal component equals, 
in the linear approximation
\begin{eqnarray}
u^{0} &=& \frac{1}{\sqrt{-g_{00}}}
={\rm e}^{-\nu/2} \left( 1-\frac{1}{2} \, r^l \, H_0 \, Y_l^m \, {\rm e}^{{\rm i} \omega t} \right), 
\label{u0}\\
\quad u^j &=& v^j \, {\rm e}^{-\nu/2},
\label{umu1}
\end{eqnarray}
where 
\begin{equation}
v^j \equiv \frac{{\rm d}x^{j}}{{\rm d}t}
\label{vj}
\end{equation}
is the $j$-th component of the velocity of the normal liquid. 
Here and below $j$
is the spatial index, $j=r$, $\theta$, and $\varphi$.
Similarly, 
using Eqs.\ (\ref{normalization}) and (\ref{uw})
one can show that for small deviations from equilibrium 
\begin{equation}
w^{0}_{(i)}=0,
\label{w}
\end{equation}
while the spatial components $w^{j}_{(i)}$ 
are small quantities, linear in perturbation
(for a similar consideration see \citealt{ga06}).
In what follows, 
instead of the four-vectors $w^{\mu}_{(i)}$ 
[which are constrained by Eq.\ (\ref{quasineutrality})]
it will be often more convenient 
to use the quantity $X^{\mu}$, 
defined by (\ref{X0}).
For small perturbations
\begin{equation}
X^0=0, 
\label{x0}
\end{equation}
while $X^j$ is non-zero but small (linear in perturbations).

Using Eqs.\ (\ref{u0}), (\ref{umu1}) and (\ref{x0}), 
as well as the definition (\ref{ub}), 
it is easy to write out an expression 
for the baryon four-velocity $U^{\mu}_{(b)}$ 
in the perturbed star,
\begin{eqnarray}
U^{0}_{({\rm b})}&=&\frac{1}{\sqrt{-g_{00}}}
={\rm e}^{-\nu/2} \left( 1-\frac{1}{2} \, r^l \, H_0 \, Y_l^m \, {\rm e}^{{\rm i} \omega t} \right), 
\label{ub0}\\
\quad U^j_{({\rm b})}&=& v^j_{({\rm b})} \, {\rm e}^{-\nu/2},
\label{ub1}
\end{eqnarray}
where the last equality 
is the definition of the $j$-th component 
of the baryon velocity $v^j_{({\rm b})}$
(linear in perturbation).
Notice that,  
as follows from Eqs.\ (\ref{ub0}) and (\ref{ub1}), 
in the linear approximation the normalization condition for the
baryon four-velocity is the same
\begin{equation}
U_{({\rm b}) \, \mu} U^{\mu}_{({\rm b})} =-1,
\label{normub}
\end{equation}
as for $u^{\mu}$ 
\footnote{
However, 
beyond the linear approximation,
Eqs.\ (\ref{X0}), (\ref{ub}), 
(\ref{normalization}), and (\ref{uw}) 
yield
$U_{({\rm b}) \, \mu} U^{\mu}_{({\rm b})}
=-1 + Y_{{\rm n}i}Y_{{\rm n}k} \, w_{(i) \,\mu}w^{\mu}_{(k)}/n_{\rm b}^2$.
The normalization condition (\ref{normub}) 
is generally not fulfilled
because the reference frame in which $U^{\mu}_{({\rm b})}=(1,0,0,0)$
is not comoving
[that is, $j^{\mu}_{({\rm b})} U_{({\rm b}) \, \mu} \neq - n_{\rm b}$ 
in that reference frame].
As it was already indicated in Sec.\ \ref{Sec:hydro}, 
all thermodynamic variables are measured in the reference frame, 
in which $u^{\mu}=(1,0,0,0)$. \label{umuumu}}.
In what follows instead of the velocities $v^j$ and $v_{({\rm b})}^j$, 
it will be more convenient to use the corresponding 
{\it Lagrangian displacements}.
They are defined by the equalities
\begin{eqnarray}
v^j \equiv \frac{\partial \xi^j}{\partial t}={\rm i} \omega \xi^j,
\label{xi}\\
v^j_{({\rm b})} \equiv \frac{\partial \xi^j_{({\rm b})} }{\partial t}= {\rm i} \omega \xi^j_{({\rm b})}.
\label{xib}
\end{eqnarray}
Introducing also the analogue 
of the Lagrangian displacement $\xi^j_{({\rm sfl})}$ 
for the vector $X^j$, 
one can write
\begin{equation}
X^j \equiv {\rm e}^{-\nu/2} \, \frac{\partial \xi^j_{({\rm sfl})}}{\partial t}
={\rm i} \omega \, {\rm e}^{-\nu/2} \, \xi_{({\rm sfl})}^j.
\label{xisfl}
\end{equation}
In terms of the Lagrangian displacements
the equality (\ref{ub}) can be presented as
\begin{equation}
\xi_{({\rm b})}^j=\xi^j+\xi_{({\rm sfl})}^j.
\label{ub2}
\end{equation}

Because of the spherical symmetry of the unperturbed star
it is sufficient to consider Lagrangian displacements
$\xi^j$, $\xi^j_{({\rm b})}$, and $\xi^j_{({\rm sfl})}$,
of the form [see also a note after Eq. (\ref{sfl2}) below]
\begin{eqnarray}
\xi^j &=& \left[\xi^r, \, \xi^{\theta}, \, \xi^{\varphi} \right]
= \left[{\rm e}^{-\lambda/2} \, r^{l-1} \, W(r) \, Y_{l}^0, \,\,\,
-r^{l-2} \, V(r) \,\, \partial_{\theta}Y_{l}^0, \,\,\, 0 \right]{\rm} {\rm e}^{{\rm i} \omega t},
\label{xi2}\\
\xi^j_{({\rm b})} &=& \left[\xi^r_{({\rm b})}, \, \xi^{\theta}_{({\rm b})}, \, \xi^{\varphi}_{({\rm b})} \right]
= \left[{\rm e}^{-\lambda/2} \, r^{l-1} \, W_{\rm b}(r) \, Y_{l}^0, \,\,\,
-r^{l-2} \, V_{\rm b}(r) \,\, \partial_{\theta}Y_{l}^0, \,\,\, 0 \right]{\rm} {\rm e}^{{\rm i} \omega t},
\label{xib2}\\
\xi^j_{({\rm sfl})} &=& \left[\xi^r_{({\rm sfl})}, \, \xi^{\theta}_{({\rm sfl})}, \, \xi^{\varphi}_{({\rm sfl})} \right]
= \left[{\rm e}^{-\lambda/2} \, r^{l-1} \, W_{\rm sfl}(r) \, Y_{l}^0, \,\,\,
-r^{l-2} \, V_{\rm sfl}(r) \,\, \partial_{\theta}Y_{l}^0, \,\,\, 0 \right]{\rm} {\rm e}^{{\rm i} \omega t}, 
\label{xisfl2}
\end{eqnarray}
where $W$, $V$, $W_{\rm b}$, $V_{\rm b}$, 
$W_{\rm sfl}$, and $V_{\rm sfl}$ 
are some functions of $r$ to be derived from oscillation equations.
In Eqs.\ (\ref{xi2})--(\ref{xisfl2}) 
$Y_{l}^0=\sqrt{(2 l+1)/(4 {\rm \pi})} \, P_{l}(\cos \theta)$, 
where $P_{l}$ is the Legendre polynomial.
Here and below we consider only spherical harmonics with $m=0$.
We can do this without any loss of generality, because, 
due to the spherical symmetry of the unperturbed star, 
oscillation eigenfrequencies as well as eigenfunctions 
$H_0$, $H_1$,$\ldots$, $W_{\rm sfl}$, and $V_{\rm sfl}$,
introduced in this section, cannot depend on $m$ 
(see, e.g., \citealt{tc67}).

It follows from Eqs.\ (\ref{ub2}) and (\ref{xi2})--(\ref{xisfl2}) 
that
\begin{eqnarray}
W_{{\rm b}} &=& W+W_{{\rm sfl}},
\label{WW}\\
V_{{\rm b}} &=& V+V_{{\rm sfl}}.
\label{V}
\end{eqnarray}
%

\section{Damping of oscillations due 
to the bulk and shear viscosities: general formulas}
\label{Sec:damping_general}
In the present paper 
among the possible mechanisms 
of dissipation of oscillation energy
we take into account
damping due to the bulk and shear viscosities
as well as due to radiation of gravitational waves.
Dissipation makes the oscillation frequency $\omega$ complex,
so that it can be presented in the form,
%
\begin{equation}
\omega=\sigma+\frac{{\rm i}}{\tau},
\label{w2}
\end{equation}
where $\sigma$ is the real part of the frequency, 
and $\tau$ is the characteristic damping time.
Assuming that damping is weak, 
in the linear approximation
one can present 
the following standard expression for $\tau$, 
\begin{equation}
\frac{1}{\tau} = 
-\frac{1}{2 E_{\rm mech}} \, \frac{{\rm d}E_{\rm mech}}{{\rm d} t},
\label{tau}
\end{equation}
where $E_{\rm mech}$ is the mechanical energy of oscillations;
${\rm d}E_{\rm mech}/{\rm d}t$ is the dissipation rate of the mechanical energy,
which can be presented as
\begin{equation}
\frac{{\rm d}E_{\rm mech}}{{\rm d} t} =
-\mathfrak{W}_{\rm bulk}-\mathfrak{W}_{\rm shear}-\mathfrak{W}_{\rm grav},
\label{Edis}
\end{equation}
where $\mathfrak{W}_{\rm bulk}$, $\mathfrak{W}_{\rm shear}$, 
and $\mathfrak{W}_{\rm grav}$ are 
the energy, dissipated per unit time
due to the bulk viscosity, shear viscosity, 
and gravitational radiation, respectively.
Introducing partial damping times 
$\tau_{\rm bulk}$, $\tau_{\rm shear}$, and $\tau_{\rm grav}$
according to
\begin{eqnarray}
\frac{1}{\tau_{\rm bulk}} &=& \frac{\mathfrak{W}_{\rm bulk}}{2 E_{\rm mech}},
\label{taubulk}\\
\frac{1}{\tau_{\rm shear}} &=& \frac{\mathfrak{W}_{\rm shear}}{2 E_{\rm mech}},
\label{taushear}\\
\frac{1}{\tau_{\rm grav}} &=& \frac{\mathfrak{W}_{\rm grav}}{2 E_{\rm mech}},
\label{taugrav}
\end{eqnarray}
one can rewrite the expression for $\tau$ as
\begin{equation}
\frac{1}{\tau}=\frac{1}{\tau_{\rm grav}}\
+\frac{1}{\tau_{\rm shear}}+\frac{1}{\tau_{\rm bulk}}.
\label{tau2}
\end{equation}
Thus, to calculate $\tau$ 
we need to know 
the mechanical energy $E_{\rm mech}$ of NS oscillations, 
as well as the quantities
$\mathfrak{W}_{\rm bulk}$, $\mathfrak{W}_{\rm shear}$, 
and $\mathfrak{W}_{\rm grav}$.

\subsection{Mechanical energy}
\label{SubSec:energy}

The general relativistic expression for the mechanical energy 
of oscillating normal (nonsuperfluid) NS 
was obtained by \cite{tc67} (see also \citealt{mt66}).
Their result can be easily generalized 
to the case of superfluid matter.
Mechanical energy $E_{\rm mech}$ 
is related to the averaged over 
the oscillation period $2{\rm \pi}/\sigma$ 
kinetic energy $\overline{E}_{\rm kin}$ 
by the standard formula,
\begin{equation}
E_{\rm mech} = 2 \, \overline{E}_{\rm kin}.
\label{Emech}
\end{equation}
Thus, to determine $E_{\rm mech}$
one needs to know $E_{\rm kin}$.
One can write (e.g., \citealt{tc67})
\begin{equation}
E_{\rm kin} = \int_{\rm star} \epsilon_{\rm kin} \, {\rm e}^{\nu/2} \, dV,
\label{Ekin}
\end{equation}
where $dV= r^2 \, {\rm e}^{\lambda/2} 
\, {\rm sin} \theta \, {\rm d}\theta \, {\rm d}\varphi \,{\rm d}r$ 
is the proper volume element;
$\epsilon_{\rm kin}$ is the kinetic energy density measured
in the locally flat coordinate system ${\tilde x}^\mu$ 
[with the metric 
$-{\rm d}s^2={\tilde g}_{\mu\nu} {\rm d} {\tilde x}^{\mu} {\rm d} {\tilde x}^{\nu}$, 
where ${\tilde g}_{\mu\nu}={\rm diag}(-1,\, 1, \,1, \,1)$],
which is at rest 
with respect to the unperturbed star.
If the NS matter is normal, $\epsilon_{\rm kin}$ is given by
\begin{equation}
\epsilon_{\rm kin} = \frac{1}{2} \, (P+\varepsilon) \,
\left[({\tilde u}^r)^2+({\tilde u}^\theta)^2 + ({\tilde u}^\varphi)^2 \right].
\label{eps_normal}
\end{equation}
Here 
\begin{equation}
{\tilde u}^j 
= \frac{\partial {\tilde x}^j}{\partial x^{\mu}} u^\mu 
\approx  {\rm e}^{-\nu/2} \, 
\left[ {\rm e}^{\lambda/2} \, v^r,\,\, r \, v^\theta, \,\, r \,{\rm sin} \theta \, v^\varphi
\right]
\label{ui_loc}
\end{equation}
is the physical velocity of the fluid 
in the locally flat coordinate system ${\tilde x}^\mu$.
For superfluid matter 
Eq.\ (\ref{eps_normal}) should be modified,
because in this case not only motion of the normal liquid component
contribute to $\epsilon_{\rm kin}$ 
but also that of the superfluid component.
Using formula (56) of \cite{kg09} one obtains
\footnote{
This expression is analogous 
to the corresponding formula for the kinetic energy density
of a nonrelativistic superfluid mixture, 
obtained by \cite{ab75}, see their equation (7).
}
%
\begin{equation}
\epsilon_{\rm kin}=\frac{1}{2}
\left\{
(P+\varepsilon) [{\tilde u}^j]^2 
+ Y_{ik} \left[ 
\mu_i \, {\tilde w}_{(k)}^j {\tilde u}_j 
+ \mu_k \, {\tilde w}_{(i)}^j {\tilde u}_j 
+ {\tilde w}_{(i)}^j{\tilde w}_{(k) \, j} \right] 
\right\},
\label{eps_sfl}
\end{equation}
where 
${\tilde w}^j_{(i)}=[\partial {\tilde x}^j/\partial x^{\mu}] \,  w^\mu_{(i)}$.
Taking into account 
Eqs.\ (\ref{quasineutrality})--(\ref{ub}) and (\ref{PPP2}),
Eq.\ (\ref{eps_sfl}) can be rewritten as
\begin{equation}
\epsilon_{\rm kin} = \frac{1}{2} \,
(P+\varepsilon) \left\{ \left[{\tilde U}_{({\rm b})}^j \right]^2 +
y \left[{\tilde X}^j \right]^2
\right\},
\label{eps_sfl2}
\end{equation}
where we neglected `temperature' term $T S$ 
in the expression (\ref{PPP2}).
In Eq.\ (\ref{eps_sfl2})  
\begin{eqnarray}
y &\equiv& \frac{n_{\rm b} Y_{\rm pp}}{\mu_{\rm n}(Y_{\rm nn}Y_{\rm pp}-Y_{\rm np}^2)}-1,
\label{y}\\
{\tilde U}_{({\rm b})}^j &=& 
\frac{\partial {\tilde x}^j}{\partial x^{\mu}} \,  U_{({\rm b})}^\mu
= {\rm e}^{-\nu/2} \, \left[ {\rm e}^{\lambda/2} \, v_{({\rm b})}^r,\,\, r \, v_{({\rm b})}^\theta, 
\,\, r \,{\rm sin} \theta \, v_{({\rm b})}^\varphi\right],
\label{Ubj}\\
{\tilde X}^j &=& \frac{\partial {\tilde x}^j}{\partial x^{\mu}} \,  X^\mu
= \left[ {\rm e}^{\lambda/2} \, X^r,\,\, r \, X^\theta, \,\, r \,{\rm sin} \theta \, X^\varphi
\right].
\label{Xj}
\end{eqnarray}
Now, using Eqs. (\ref{xib}), (\ref{xisfl}), (\ref{xib2}), and (\ref{xisfl2})
let us express (\ref{Ubj}) and (\ref{Xj}) 
through the functions $W_{\rm b}(r)$, $V_{\rm b}(r)$, 
$W_{\rm sfl}(r)$, and $V_{\rm sfl}(r)$, 
and then substitute Eq.\ (\ref{eps_sfl2}) for $\epsilon_{\rm kin}$ 
into (\ref{Ekin}). 
After integrating Eq.\ (\ref{Ekin}) 
over ${\rm sin}\theta \, {\rm d}\theta \, {\rm d} \varphi$
(in the same way as it was done in \citealt{tc67})
and making use of Eq.\ (\ref{Emech}), 
one arrives at the following 
expression for $E_{\rm mech}$,
\begin{equation}
E_{\rm mech}(t)=E_{\rm mech \, (b)}(t)+E_{\rm mech \, ( sfl)}(t),
\label{Emech2}
\end{equation}
where we tentatively presented $E_{\rm mech}$
as a sum of two terms related 
to the baryon motion as a whole
$E_{\rm mech \, (b)}$ 
and an additional term $E_{\rm mech \, (sfl)}$
appearing because of the superfluid motion,
\begin{eqnarray}
E_{\rm mech \, (b)}(t) &=& \frac{1}{2} \, \sigma^2 \, {\rm e}^{-2t/\tau} 
\int_0^R (P+\varepsilon) \, e^{(\lambda-\nu)/2} \, r^{2l} \left[
W_{\rm b}^2  
+ l (l+1) \,V_{\rm b}^2
\right]  {\rm d}r,
\label{Emechb}\\
E_{\rm mech \, (sfl)}(t) &=& \frac{1}{2} \, \sigma^2 \, {\rm e}^{-2t/\tau} 
\int_0^R (P+\varepsilon) \, e^{(\lambda-\nu)/2} \, r^{2l} \, y \left[
 W_{\rm sfl}^2
+ l (l+1)  \, V_{\rm sfl}^2
\right] {\rm d}r.
\label{Emechsfl}
\end{eqnarray}
Strictly speaking, 
the functions $W_{\rm b}(r)$, $V_{\rm b}(r)$, 
$W_{\rm sfl}(r)$, and $V_{\rm sfl}(r)$ 
in these formulas 
are complex, 
that is, instead of, 
for example, $W_{\rm b}(r)^2$
one should write $|W_{\rm b}(r)|^2$.
Notice, however, that all these functions
[as well as $H_0(r)$, $H_1(r)$, $H_2(r)$, and $K(r)$] 
are defined up to {\it the same}
arbitrary complex multiplicative constant. 
Since $\sigma \gg 1/\tau$ 
(dissipation is weak),
one can always choose the constant in such a way, 
that the real parts of {\it all} these functions 
would be much greater than their imaginary parts
(e.g., ${\rm Re}[H_{2}(r)]\gg {\rm Im}[H_{2}(r)]$),
so that one could neglect their `complexity'. 
From here on, unless otherwise stated,
by the functions $W_{\rm b}(r)$, $V_{\rm b}(r)$, 
$W_{\rm sfl}(r)$, $V_{\rm sfl}(r)$,  
$H_0(r)$, $H_1(r)$, $H_2(r)$, and $K(r)$
we mean their real parts.

In the absence of superfluidity 
$W_{\rm sfl}=V_{\rm sfl}=0$, 
$W_{\rm b}=W$, and $V_{\rm b}=V$. 
In that case Eq.\ (\ref{Emechb}) 
gives a mechanical energy of a nonsuperfluid star
that coincides, up to notations,
with the corresponding expression (29) of \cite{tc67}. 
%

\subsection{Dissipation rates}
\label{SubSec:rates}

The damping time $\tau_{\rm grav}$ 
due to radiation of gravitational waves 
can be obtained from the equations,
describing linear oscillations of NSs 
(see Sec.\ \ref{SubSec:osc} below). 
The goal of the present section is to determine 
the dissipation rate of oscillation energy
due to the bulk $\mathfrak{W}_{\rm bulk}$ 
and shear $\mathfrak{W}_{\rm shear}$ viscosities 
and, as a consequence, 
the damping times 
$\tau_{\rm bulk}$ and $\tau_{\rm shear}$.

For that, 
we turn to the entropy generation equation (\ref{entropy3}).
Using it, one can find rate of change of the
(averaged over the oscillation period) 
thermal energy of a star ${\rm d}E_{\rm th}/{\rm d}t$  
due to bulk and shear viscosities.
Following the derivation of Eq. (34) in \cite*{gyg05},
one obtains
\begin{equation}
\frac{{\rm d}E_{\rm th}}{{\rm d}t}= \int_{\rm star} (\overline{Q}_{\rm bulk} 
+ \overline{Q}_{\rm shear}) \, {\rm e}^{\nu} \, {\rm d}V,
\label{Eth}
\end{equation}
where $\overline{Q}_{\rm bulk}$ and $\overline{Q}_{\rm shear}$ 
are the values of $Q_{\rm bulk}$ and $Q_{\rm shear}$, 
averaged over the oscillation period $2 {\rm \pi}/\sigma$ 
[see Eqs.\ (\ref{bulk}) and (\ref{shear})].

Obviously, the increase in the thermal energy $E_{\rm th}$ 
is accompanied by the decrease of the oscillation energy $E_{\rm mech}$, 
that is
\begin{eqnarray}
\mathfrak{W}_{\rm bulk}=\int_{\rm star} \overline{Q}_{\rm bulk} \, {\rm e}^{\nu} \, {\rm d}V,
\label{Wbulk}\\
\mathfrak{W}_{\rm shear}=\int_{\rm star} \overline{Q}_{\rm shear} \, {\rm e}^{\nu} \, {\rm d}V.
\label{Wshear}
\end{eqnarray}
Using these equations, 
as well as the formulas (\ref{bulk}), (\ref{shear}),
(\ref{taubulk}), (\ref{taushear}), 
and the definitions of Sec.\ \ref{SubSec:small},
one gets, after rather lengthy calculations,
\begin{eqnarray}
\frac{1}{\tau_{\rm bulk}}&=&
\frac{\sigma^2}{4 E_{\rm mech}(0)} \,
\int_{0}^{R} \, r^{2(l+1)}  \, {\rm e}^{\lambda/2}\,
\left[
\sqrt{\xi_2} \, \beta_1 + \sqrt{\xi_{3{\rm n}}} \, \beta_2
\right]^2 \, {\rm d}r,
\label{taubulk2}\\
\frac{1}{\tau_{\rm shear}} &=&
\frac{\sigma^2}{2 E_{\rm mech}(0)} \,
\int_{0}^{R} \, \eta \, r^{2(l-1)} \, {\rm e}^{\lambda/2} \, 
\nonumber\\
&& \times \left\{
\frac{3}{2} \, (\alpha_1)^2 + 2l (l+1) \,(\alpha_2)^2
+ l (l+1) \,
    \left[ \frac{1}{2} \, l (l+1) -1
  \right]  V^2
\right\}  {\rm d}r,
\label{taushear2}
\end{eqnarray}
where
\begin{eqnarray}
\beta_1(r) &=& K+\frac{1}{2} \, H_2 -
\frac{1}{r} \, {\rm e}^{-\lambda/2} 
\left[ 
\frac{{\rm d} W}{{\rm d}r}+\frac{1}{r}(l+1)\, W
\right]
-l(l+1) \, \frac{V}{r^2},
\label{Pxi2}\\
\beta_2(r) &=& -\frac{1}{r} \, {\rm e}^{-\lambda/2} 
\left[ 
\frac{{\rm d} (n_{\rm b} \, W_{\rm sfl})}{{\rm d}r}+\frac{1}{r}(l+1)\, n_{\rm b}\, W_{\rm sfl}
\right]
-l(l+1) \, \frac{n_{\rm b}\, V_{\rm sfl}}{r^2},
\label{Pxi3}\\
\alpha_1(r) &=& \frac{r^2}{3} \, 
\left\{
\frac{2}{r}\, {\rm e}^{-\lambda/2} \, \left[ \frac{{\rm d}W}{{\rm d}r}+(l-2) \frac{W}{r}
\right]
+K-H_2 - l(l+1) \, \frac{V}{r^2}
\right\},
\label{alpha1}\\
\alpha_2(r) &=&\frac{r}{2} \,
\left[
\frac{{\rm d}V}{{\rm d}r}+(l-2) \, \frac{V}{r}-{\rm e}^{\lambda/2}\, \frac{W}{r}
\right]  {\rm e}^{-\lambda/2}.
\label{alpha2}
\end{eqnarray}
As for the mechanical energy (\ref{Emech2}),
to obtain from these formulas 
$\tau_{\rm bulk}$ and $\tau_{\rm shear}$ for a nonsuperfluid star, 
one has to put $W_{\rm sfl}=V_{\rm sfl}=0$.
In that case our Eqs.\ (\ref{taubulk2}) and (\ref{taushear2})
should coincide with the corresponding formulas (5) and (6) of \cite{cls90}. 
Unfortunately, direct comparison of these formulas reveals, 
that our $\tau_{\rm bulk}$ and $\tau_{\rm shear}$ 
appear to be 2 times larger. 
Using, as tests examples, damping of:
($i$) NS radial oscillations,
($ii$) $p$-modes in the NS envelopes,
and ($iii$) sound waves in the nonsuperfluid matter of NSs
we checked, 
that our results reproduce those 
of \cite{gyg05, cy05, kg09},
obtained
in a quite a different way.

\section{Oscillation equations}
\label{SubSec:osc}

In order to calculate 
$\tau_{\rm bulk}$ and $\tau_{\rm shear}$ 
one has to determine 
the oscillation eigenfrequencies $\sigma$ 
and eigenfunctions 
$H_0$, $H_1$, $H_2$, $K$, $W_{\rm b}$, $V_{\rm b}$, 
$W_{\rm sfl}$, and $V_{\rm sfl}$.
To do that one needs to formulate oscillation equations.
Since the dissipation is weak, 
when deriving the oscillation equations
one can neglect the dissipative terms 
in the superfluid hydrodynamics 
of Sec.\ \ref{Sec:hydro}
and put $\tau^{\mu\nu}=0$ and $\varkappa_{\rm n}=0$. 

As it was shown in \cite{gk11}, equations, 
describing small linear oscillations of a NS include:

($i$) Continuity equations for baryons (\ref{cont_b}) 
and electrons (\ref{cont_e}),
that can be written 
in terms of the baryon and electron 
number density perturbations, 
$\delta n_{\rm b}$ and $\delta n_{\rm e}$, as 
\begin{eqnarray}
\delta n_{\rm b} &=& \frac{{\rm i}}{\omega \, {\rm e}^{-\nu/2}} 
\left[ 
\partial_j (n_{\rm b}) \, U_{\rm (b)}^j + n_{\rm b} \, U^\mu_{\rm (b)\, ; \mu}
\right],
\label{dnb}\\
\delta n_{\rm e} &=& \delta n_{\rm e \, ({\rm norm})} 
+ \delta n_{\rm e \, ({\rm sfl})},
\label{dne}
\end{eqnarray}
where $j$ is the spatial index and we defined
\begin{eqnarray}
\delta n_{\rm e \, ({\rm norm})} &\equiv&
\frac{{\rm i}}{\omega \, {\rm e}^{-\nu/2}} 
\left[ 
\partial_j (n_{\rm e}) \, U_{\rm (b)}^j + n_{\rm e} \, U^\mu_{\rm (b)\, ; \mu}
\right],
\label{dnenorm}\\
\delta n_{\rm e \, ({\rm sfl})} &\equiv&
-\frac{{\rm i}}{\omega \, {\rm e}^{-\nu/2}} 
\left[ 
\partial_j (n_{\rm e}) \, X^j + n_{\rm e} \, X^\mu_{; \mu}
\right].
\label{dnesfl}
\end{eqnarray}

($ii$) Einstein equations, 
which can schematically be presented as 
\begin{equation}
\delta(R^{\mu \nu} - 1/2 \,\, g^{\mu \nu} \, R) 
=8 {\rm \pi} G\, \, \delta T^{\mu \nu},
\label{einst}
\end{equation}
where the perturbation $\delta T^{\mu \nu}$ 
of the energy-momentum tensor (\ref{Tmunu})
can be expressed in terms 
of the perturbations 
of baryon four-velocity $\delta U^{\mu}_{({\rm b})}$, 
metric $\delta g_{\mu\nu}$, 
pressure $\delta P$ 
and energy density $\delta \varepsilon$ as
\begin{equation}
\delta T^{\mu \nu} = 
(\delta P + \delta \varepsilon) \, U^{\mu}_{({\rm b})}  U^{\nu}_{({\rm b})} 
+ 
(P + \varepsilon ) 
\left[U^{\mu}_{({\rm b})} \, \delta U^{\nu}_{({\rm b})}+ U^{\nu}_{({\rm b})} 
\, \delta U^{\mu}_{({\rm b})} \right] + \delta P \, g^{\mu \nu} 
+ P \, \delta g^{\mu \nu}. 
\label{deltaTmunu}
\end{equation}
In Eq.\ (\ref{einst}) $R^{\mu\nu}$ and $R$
are the Ricci tensor and scalar curvature, respectively;
$G$ is the gravitation constant.

($iii$) `Superfluid' equation, 
that can be derived 
from Eqs.\ (\ref{Tmunu_cons}) and (\ref{wmu22}) 
of Sec.\ \ref{Sec:hydro} 
(here we present only the spatial components $j$ of this equation) 
\footnote{
It is worth to make a number of comments on Eq. (\ref{sfl}):
($i$) In \cite{gk11} this equation was derived under the assumption 
that the only superfluid species are neutrons
(that is $Y_{{\rm p}i}=0$). 
A generalization of this equation to the case of 
possible proton superfluidity is presented in \cite{cg11} 
[see their Eq. (3)];
($ii$) In both papers, \cite{gk11, cg11}, 
this equation is written with the same mistake.
In particular, 
in \cite{cg11}
one should write
$n_{\rm e} \,\, \partial_j ({\rm e}^{\nu/2} \,\delta \mu)$
instead of
$n_{\rm e} \,{\rm e}^{\nu/2} \,\, \partial_j (\delta \mu)$
in the right-hand side of Eq. (3).}
%
\begin{equation}
{\rm i} \, \omega \, 
(\mu_{\rm n} \, Y_{{\rm n}k} \, w_{(k)j} - n_{\rm b} \, w_{({\rm n})j})
= n_{\rm e} \,\, \partial_j ({\rm e}^{\nu/2} \,\delta \mu).
\label{sfl}
\end{equation}
Expressing the vectors $w^{j}_{(i)}$ through $X^{j}$ 
in this equation
[see Eqs.\ (\ref{quasineutrality}) and (\ref{X0})], 
and introducing the redshifted imbalance of chemical potentials
$\delta \mu^{\infty} \equiv {\rm e}^{\nu/2}\, \delta \mu$,
one can rewrite Eq.\ (\ref{sfl}) as
\begin{equation}
X_j 
= \frac{{\rm i} \, n_{\rm e}}{\mu_{\rm n} n_{\rm b} \, \omega \, y} 
\,\partial_j (\delta \mu^\infty),
\label{sfl2}
\end{equation}
where $y$ is defined by Eq.\ (\ref{y}). 
Notice, that this equation dictates
the most general form 
of the superfluid Lagrangian displacement 
$\xi^j_{\rm (sfl)}$,
that was already obtained 
in Eq.\ (\ref{xisfl2}) from the symmetry arguments.

Eqs. (\ref{dnb})--(\ref{sfl2}) 
should be supplemented 
with the expressions for the perturbations $\delta P$,
$\delta \mu$, and $\delta \varepsilon$.
To derive them, let us notice that
any thermodynamic quantity (e.g., $P$)
in the superfluid matter 
can be presented as a function of
$n_{\rm b}$, $n_{\rm e}$, $T$, and $w_{(i) \, \mu} w^{\mu}_{(k)}$ 
(see, e.g., \citealt{gusakov07}). 
In strongly degenerate matter 
the dependence of $P$, $\delta \mu$, and $\varepsilon$
on $T$ can be neglected 
(see, e.g., \citealt{reisenegger95, gyg05}), 
while the scalars
$w_{(i) \, \mu} w^{\mu}_{(k)}$ 
are quadratically small
in a slightly perturbed star 
[see Sec.\ \ref{SubSec:small}].
Thus, $P=P(n_{\rm b}, \, n_{\rm e})$, 
$\delta \mu=\delta \mu(n_{\rm b}, \, n_{\rm e})$, 
and $\varepsilon=\varepsilon(n_{\rm b}, \, n_{\rm e})$.
Expanding these functions 
into Taylor series near the equilibrium, one obtains
\begin{eqnarray}
\delta P &=& n_{\rm b} \, \frac{\partial P}{\partial n_{\rm b}}
\left[
\frac{\delta n_{\rm b}}{n_{\rm b}} + {\tilde s} \, \frac{\delta n_{\rm e \, ({\rm norm})}}{n_{\rm e}}
+ s \, \frac{\delta n_{\rm e \, ({\rm sfl})}}{n_{\rm e}} 
\right],
\label{P}\\
\delta \mu &=& n_{\rm e} \, \frac{\partial \delta\mu}{\partial n_{\rm e}}
\left[
z \, \frac{\delta n_{\rm b}}{n_{\rm b}} + \frac{\delta n_{\rm e \, ({\rm norm})}}{n_{\rm e}}
+ \frac{\delta n_{\rm e \, ({\rm sfl})}}{n_{\rm e}} 
\right],
\label{dmu3}\\
\delta\varepsilon &=& \mu_{\rm n} \, \delta n_{\rm b}.
\label{e}
\end{eqnarray}
where we made use of Eq.\ (\ref{dne}), 
and introduced dimensionless {\it coupling parameter} $s$ 
and the quantities $\tilde{s}$ and $z$,
\begin{eqnarray}
s &\equiv& \frac{n_{\rm e}}{n_{\rm b}} \, 
\frac{(\partial P/\partial n_{\rm e})}{(\partial P/\partial n_{\rm b})},
\label{s}\\
\tilde{s} &\equiv& \frac{n_{\rm e}}{n_{\rm b}} \, 
\frac{(\partial P/\partial n_{\rm e})}{(\partial P/\partial n_{\rm b})},
\label{stilde}\\
z &\equiv&\frac{n_{\rm b}}{n_{\rm e}} \, 
\frac{(\partial \delta \mu/\partial n_{\rm b})}{(\partial \delta \mu/\partial n_{\rm e})}.
\label{z}
\end{eqnarray}
Notice that 
the variable ${\tilde s}$
is equal to $s$ here.
The reason for
discriminating between ${\tilde s}$ and $s$
is purely technical:
To solve oscillation equations 
(see Secs.\ \ref{Sec:approach} and \ref{Sec:Results})
it turns out to be convenient to develop a perturbation theory 
in (small) parameter $s$, 
at the same time treating the terms depending on ${\tilde s}$ 
in a {\it non-perturbative} way
(see Sec.\ \ref{SubSec:strategy},
and, in particular, footnote $^{\ref{9}}$ there).
 %
%
When deriving Eq.\ (\ref{e}) 
we took into account that 
$[\partial \varepsilon(n_{\rm b}, \, n_{\rm e})/\partial n_{\rm e}] \, 
\delta n_{\rm e}=-\delta \mu \, \delta n_{\rm e}$ 
is a quadratically small quantity, 
because $\delta \mu=0$ in equilibrium 
\footnote{
The equality  
$\partial \varepsilon(n_{\rm b}, \, n_{\rm e})/\partial n_{\rm e} =- \delta \mu$ 
follows from the second law of thermodynamics (\ref{2ndlaw}), 
which can be rewritten in our case as 
${\rm d} \varepsilon= \mu_{\rm n} \, {\rm d} n_{\rm b}-\delta \mu \, {\rm d} n_{\rm e}$.}.

The vector superfluid equation (\ref{sfl2}) 
can be substantially simplified, 
and reduced to a scalar one. 
For that let us notice that, 
without any loss of generality, 
the scalar $\delta \mu^\infty$ can be presented as
\begin{equation}
\delta \mu^\infty= \delta \mu_{l}(r) \, Y_l^0(\theta).
\label{muinfty}
\end{equation}
Employing now Eqs.\ (\ref{sfl2}) and (\ref{dmu3}), 
one arrives at
\begin{equation}
    \delta \mu_l^{\prime\prime}+\left(\frac{h^\prime}{h}
-\frac{\lambda^\prime}{2}
    +\frac{2}{r}\right)\delta
    \mu_l^\prime 
-{\rm e}^\lambda\left[\frac{l(l+1)}{r^2}+\mathrm
    e^{-\nu/2}\frac{\omega^2}{h \, \mathfrak{B}}
    \right] \delta \mu_l =
    -\frac{\omega^2 \, {\rm e}^{\lambda-\nu/2}}{h \mathfrak{B}} \,\delta \mu_{{\rm norm}  \, l}.
\label{sfl3}
\end{equation}
Here $h={\rm e}^{\nu/2} \,n_{\rm e}^2/(\mu_{\rm n} \,n_{\rm b} \,y)$, 
$\mathfrak{B} \equiv \partial \delta \mu(n_{\mathrm b}, n_{\mathrm e})/\partial n_\mathrm e$,
and prime ($\prime$) means derivative 
with respect to the radial coordinate $r$.
Furthermore, $\delta \mu_{{\rm norm}  \, l}(r)$ 
in Eq.\ (\ref{sfl3}) 
is defined by
\begin{equation}
\delta \mu_{\rm norm}^\infty
= \delta\mu_{{\rm norm}  \, l}(r) \, Y_l^0(\theta),
\label{dmu5}
\end{equation}
where
\begin{equation}
\delta \mu_{\rm norm}^\infty \equiv {\rm e}^{\nu/2} \,
n_{\rm e} \, 
\mathfrak{B}
\left[
z \, \frac{\delta n_{\rm b}}{n_{\rm b}} + \frac{\delta n_{\rm e \, ({\rm norm})}}{n_{\rm e}} 
\right]
\label{dmunorm}
\end{equation}
is a part of $\delta \mu^\infty$, 
which depends on $\delta g^{\mu\nu}$ and $U^{\mu}_{({\rm b})}$ 
and is independent of the superfluid degrees of freedom $X^j$ 
[see Eqs.\ (\ref{dnb}) and (\ref{dnenorm})].
The function $\delta \mu_{{\rm norm}\, l}(r)$ 
can be easily rewritten in terms of
$H_{0}(r)$, $H_1(r)$, $H_{2}(r)$, $K(r)$, 
$W_{\rm b}(r)$, and $V_{\rm b}(r)$ 
with the help of 
Eqs.\ (\ref{metric2}), (\ref{ub0}), (\ref{ub1}), (\ref{xib}), (\ref{xib2}),
(\ref{dnb}), and (\ref{dnenorm}).
One obtains
\begin{equation}
\delta \mu_{{\rm norm}\,l} = \mathrm{e}^{\nu/2}\, 
n_\mathrm{b}
\,\frac{\partial \delta \mu (n_\mathrm{b},x_\mathrm{e})}
{\partial n_\mathrm{b}} \, 
r^l \, \beta_1,
\label{dmunorm_l}
\end{equation}
where $x_{\rm e} \equiv n_{\rm e}/n_{\rm b}$ and 
$\beta_1(r)$ is given by Eq.\ (\ref{Pxi2}) with
$W_{\rm b}$ and $V_{\rm b}$ instead of, 
respectively, $W$ and $V$.

Finally, let us mention one important property, 
that follows from the oscillation equations 
and quasineutrality condition (\ref{quasineutrality}).
If neutrons in a nonrotating nonmagnetized star are normal 
(i.e. $Y_{\rm nn}=Y_{\rm np}=0$), 
while protons are superfluid ($Y_{\rm pp}\neq 0$), 
then oscillation eigenfrequencies 
and eigenfunctions for such star
will be indistinguishable 
from that for a normal star of the same mass 
(where both protons and neutrons are nonsuperfluid).

\section{Our approach}
\label{Sec:approach}

\subsection{Decoupling of superfluid and normal modes}
\label{SubSec:dec}
%
In principle, 
Eqs.\ (\ref{dnb})--(\ref{dmunorm_l}) 
allow one to study the nonradial oscillations of superfluid NSs
and thus to determine 
the spectrum of eigenfrequencies $\omega$, 
eigenfunctions $H_0$, $H_1$,$\ldots$, $W_{\rm sfl}$, and $V_{\rm sfl}$, 
and hence the damping times 
$\tau_{\rm grav}$, $\tau_{\rm bulk}$, and $\tau_{\rm shear}$.
However, this task can be significantly simplified,
if one notes that 
the dimensionless coupling parameter $s$ (\ref{s}) 
is small for realistic equations of state 
of superdense matter (\citealt{gk11}).
For example, 
for the equation of state APR (\citealt*{apr98}) employed below
$s\sim 0.01\div0.05$. 
This means that one can look for
the solution to the system of 
Eqs.\ (\ref{dnb})--(\ref{dmunorm_l}) 
in the form of a series in $s$.
Since $s$ is small, 
the approximation $s=0$ 
is already quite accurate. 
Indeed, as it was shown in \cite{gk11} 
with the example of radial oscillations,
the eigenfrequencies calculated
in this approximation 
differ from the exact ones, 
on average, by
$\sim 1.5 \div 2 \%$.
Thus, in what follows 
all calculations are performed assuming $s=0$.

How this approach simplifies the problem?
As it was first demonstrated in \cite{gk11}, 
in the $s=0$ approximation 
superfluid degrees of freedom (vectors $X^j$)
completely decouple 
from the `normal' degrees of freedom 
[metric perturbations $\delta g_{\mu\nu}$ 
and baryon four-velocities $\delta U^{\mu}_{({\rm b})}$].
That is, one has {\it two} distinct classes of oscillations:
`superfluid' and `normal' modes,
which are described by independent equations.
For superfluid-type oscillations 
the metric and baryon velocity are not perturbed 
[$\delta g_{\mu\nu}=0$ and $\delta U^{\mu}_{({\rm b})}=0$],
hence these modes {\it do not emit} gravitational waves;
moreover, they are {\it entirely} localized in the SFL-region.
At the same time, 
the frequencies of normal modes 
are {\it indistinguishable} 
from those of a normal (nonsuperfluid) star
of the same mass
\footnote{
Here and below by `normal modes' 
we mean oscillation modes (of approximate solution), 
that also exist in the normal (nonsuperfluid) star.}.
%
Below we discuss in more detail 
decoupling of superfluid and normal oscillation modes and
how this property can be used to
calculate the characteristic damping times.

\subsection{A strategy to calculate the damping times}
\label{SubSec:strategy}

So, let us formally assume that $s=0$ 
(while ${\tilde s}$ is given 
by Eq.\ (\ref{stilde}) and is non-zero).
%
Then, as follows from Eq.\ (\ref{P}), 
$\delta P$ equals
\begin{equation}
\delta P = n_{\rm b} \, \frac{\partial P}{\partial n_{\rm b}}
\left[
\frac{\delta n_{\rm b}}{n_{\rm b}} + {\tilde s} \, \frac{\delta n_{\rm e \, ({\rm norm})}}{n_{\rm e}} 
\right]
\label{P2}
\end{equation}
%
and is independent of the superfluid degrees of freedom $X^\mu$
[see Eqs.\ (\ref{dnb}) and (\ref{dnenorm})].
Other terms in the expression (\ref{deltaTmunu}) for $\delta T^{\mu \nu}$ 
also do not depend on $X^\mu$
[in particular, 
$\delta \varepsilon=\mu_{\rm n} \delta n_{\rm b}$ 
does not depend on $X^{\mu}$ due to Eqs.\ (\ref{dnb}) and (\ref{e})].
Thus, we come to conclusion that 
the linearized Einstein equations (\ref{einst}) 
depend {\it only} on perturbations of the metric $g_{\mu\nu}$ 
and the baryon four-velocity $U^\mu_{({\rm b})}$ 
and are independent of $X^{\mu}$.
Moreover, it is easy to see, 
that in the case $s=0$ 
these equations 
(and the corresponding boundary conditions)
have exactly the same form as in the absence of superfluidity 
%
\footnote{
This is the main advantage of treating $\tilde s$ in a non-perturbative way.
Notice, however, that this trick  
leads to somewhat `excessive' 
accuracy of the approximate solution to oscillation equations:
the retained terms depending on $\tilde{s}$ 
may lead to smaller correction to the solution 
than the $s$-dependent terms which were ignored.
Bearing this in mind and with the aim
to simplify consideration, 
in \cite{gk11} it was assumed that 
both parameters $s$ and ${\tilde s}$ vanish in the s=0 approximation.
Such an approach is also possible.
In that case, strictly speaking, 
the resulting Einstein equations would differ slightly 
from the equations 
describing oscillations of a nonsuperfluid NSs.
In particular, 
instead of the standard adiabatic index
of the `frozen' npe-matter 
$\gamma_{\rm fr}= (n_{\rm b}/P)[\partial P(n_{\rm b}, x_{\rm e})/\partial n_{\rm b}]$
the new index would appear, 
$\gamma=(n_{\rm b}/P)[\partial P(n_{\rm b}, n_{\rm e})/\partial n_{\rm b}]$.
However, this difference is not essential, 
because $s$ 
and $\tilde{s}$
are small.\label{9}}.
%
Correspondingly, 
two alternatives are possible
when solving 
the system 
of Eqs.\ (\ref{dnb})--(\ref{dmunorm_l})
in the approximation $s=0$:

(1) A star oscillates at a frequency 
which is {\it not} an eigenfrequency 
of the Einstein equations (\ref{einst}).
In that case, to satisfy Eq.\ (\ref{einst}), 
one has to demand 
\begin{equation}
H_0=H_1=H_2=K=W_{\rm b}=V_{\rm b}=0.
\label{1}
\end{equation}
From Eq. (\ref{dmunorm_l}) 
it follows then, 
that $\delta \mu_{{\rm norm}\, l}=0$
and the superfluid equation (\ref{sfl3}) decouples from
the Einstein equations. 
As a result we arrive at 
the `source-free' equation 
(with the right-hand side vanished), 
first derived in \cite{cg11}
\footnote{
The corresponding equation (5) of \cite{cg11}, 
contains a mistake,
that was corrected in the second version 
of the manuscript in arXiv 
(see arXiv:1107.4242v2).},
%
\begin{equation}
    \delta \mu_l^{\prime\prime}+\left(\frac{h^\prime}{h}
-\frac{\lambda^\prime}{2}
    +\frac{2}{r}\right)\delta
    \mu_l^\prime 
-{\rm e}^\lambda\left[\frac{l(l+1)}{r^2}+\mathrm
    e^{-\nu/2}\frac{\omega^2}{h \, \mathfrak{B}}
    \right] \delta \mu_l =0.
\label{sfl4}
\end{equation}
%
This equation describes {\it superfluid modes}
and should be solved in the stellar region 
where neutrons are superfluid (SFL-region). 
It should be supplemented with a number of boundary conditions,
discussed in \cite{cg11} 
[similar, but more general boundary conditions for Eq. (\ref{sfl3}) 
are presented in the Appendix].
Having solved Eq.\ (\ref{sfl4}) for $\delta \mu_l$,
it is easy to determine 
the functions $W_{\rm sfl}$ and $V_{\rm sfl}$
using Eqs.\ (\ref{xisfl}), (\ref{xisfl2}), and (\ref{sfl2}).
Using $W_{\rm sfl}$ and $V_{\rm sfl}$, 
one can find the functions $W$ and $V$ 
from Eqs.\ (\ref{WW}), (\ref{V}), and (\ref{1}), 
\begin{equation}
W=-W_{\rm sfl}, \quad V=-V_{\rm sfl}.
\label{WV}
\end{equation}
This information is sufficient 
to calculate $\tau_{\rm bulk}$ and $\tau_{\rm shear}$ 
from Eqs.\ (\ref{taubulk2}) and (\ref{taushear2}) 
[as follows from Eq.\ (\ref{1}), 
$\tau_{\rm grav}=\infty$ 
for superfluid modes 
in the $s=0$ approximation].

(2) A star oscillates at a frequency
which is an eigenfrequency of Einstein equations (\ref{einst}).
In that case, 
the eigenfrequency and eigenfunctions 
$H_0$, $H_1$, $H_2$, $K$, $W_{\rm b}$, and $V_{\rm b}$ 
are {\it indistinguishable} from 
the corresponding eigenfrequency and eigenfunctions
for an oscillating nonsuperfluid NS
[we recall, that for the nonsuperfluid star 
$W_{\rm b}=W$, $V_{\rm b}=V$, 
because $W_{\rm sfl}=V_{\rm sfl}=0$, 
see Eqs.\ (\ref{WW}) and (\ref{V})].
There is, however, 
one very important difference: 
for a superfluid star the functions
$W_{\rm sfl}$ and $V_{\rm sfl}$
{\it do not vanish}
in the SFL-region 
and are comparable there to 
$W_{\rm b}$ and $V_{\rm b}$.
As follows from Eqs.\ (\ref{taubulk2}) and (\ref{taushear2}),
the damping times $\tau_{\rm bulk}$ and $\tau_{\rm shear}$ 
depend on these functions 
[as well as on $W=W_{\rm b}-W_{\rm sfl}$ and $V=V_{\rm b}-V_{\rm sfl}$],
that is why 
the determination of 
$W_{\rm sfl}$ and $V_{\rm sfl}$
is a necessary task.

To determine these functions 
we make use of Eq.\ (\ref{sfl3}).
Since the oscillation frequency 
$\omega=\sigma+ {\rm i}/\tau_{\rm grav}$ 
and the eigenfunctions $H_0$, $H_1$, $H_2$, $K$, $W_{\rm b}$, and $V_{\rm b}$
are already known, 
we can, 
using Eq.\ (\ref{dmunorm_l}), 
calculate $\delta \mu_{{\rm norm}\, l}$ and determine
a `source' in the right-hand side of Eq.\ (\ref{sfl3}).
This source plays a role of an external driving force, 
that makes the superfluid equation (\ref{sfl3})
`oscillate' at the frequency $\omega$, 
which is {\it not} an eigenfrequency for this equation 
\footnote{
In the present paper,  
in all numerical calculations 
we used $\sigma$
instead of $\omega$ in Eq.\ (\ref{sfl3}), 
because $\sigma\gg 1/\tau_{\rm grav}$.
Also, when calculating $\delta \mu_{{\rm norm}\, l}$
we only employed the real parts of eigenfunctions
$H_0$, $H_1$, $H_2$, $K$, $W_{\rm b}$, and $V_{\rm b}$
[see a note after Eq.\ (\ref{Emechsfl})].}. 
As a result, the function $\delta \mu_l(r)$ will be nonzero.
To determine it one has to specify 
the boundary conditions for Eq. (\ref{sfl3});
they are formulated in Appendix. 
Having solved Eq.\ (\ref{sfl3}) numerically 
and having defined 
$\delta \mu_l(r)$, 
one can calculate the functions $W_{\rm sfl}$ and $V_{\rm sfl}$,
using Eqs.\ (\ref{xisfl}), (\ref{xisfl2}), (\ref{sfl2}), and (\ref{muinfty}).

Summarizing, in the approximation $s=0$ 
the eigenfrequencies and eigenfunctions 
$H_0$, $H_1$, $H_2$, $K$, $W_{\rm b}$, and $V_{\rm b}$, 
(and hence $\tau_{\rm grav}$) for the normal modes
appear to be {\it the same} as for a nonsuperfluid star.
At the same time the eigenfunctions 
$W_{\rm sfl}$ and $V_{\rm sfl}$ are non-zero in the SFL-region
and should be determined from Eq.\ (\ref{sfl3}). 
As a result, the damping times 
$\tau_{\rm bulk}$ and $\tau_{\rm shear}$,
defined by Eqs.\ (\ref{taubulk2}) and (\ref{taushear2}),
will {\it differ} from the corresponding times, 
calculated using the ordinary (nonsuperfluid) hydrodynamics
(even if one takes into account the effects 
of superfluidity on the kinetic coefficients).

\section{Results}
\label{Sec:Results}

Let us apply the approach, 
suggested in the previous section, 
to determine the frequency spectrum and damping 
times for an oscillating superfluid NS. 
But first let us discuss 
its equilibrium model. 

\subsection{Microphysics input and equilibrium model}
\label{SubSec:micro}

%
\begin{figure}
    \begin{center}
        \leavevmode
        \epsfxsize=6.4in \epsfbox{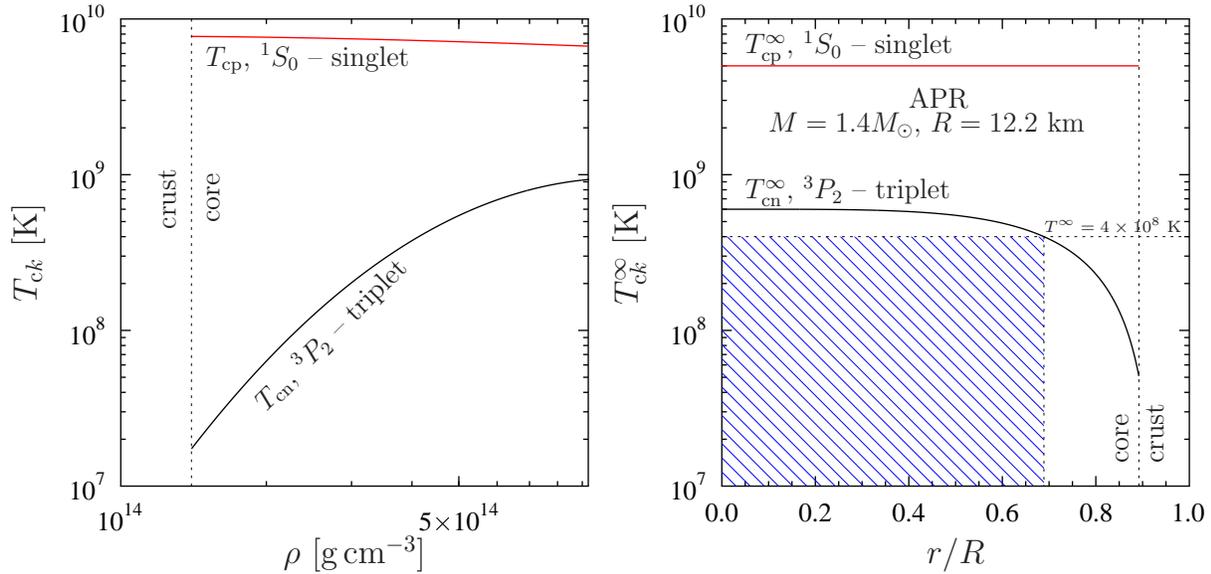}
    \end{center}
    \caption{(color online)
    Left panel: Nucleon critical temperatures
    $T_{\mathrm c k}$ ($k={\rm n}$, ${\rm p}$)
    versus density $\rho$ for model 1.
    Right panel: Redshifted critical temperatures
    $T^\infty_{\mathrm c k}$ versus radial coordinate $r$
    (in units of $R$) for model 1.
    }
    \label{Fig:Tc1}
\end{figure}

As mentioned in Sec.\ \ref{Sec:hydro}, 
we consider the simplest npe-composition of NS core. 
We adopt APR equation of state
(\citealt{apr98}) parametrized by \cite{hh99} in the core and  
the equation of state by \cite{nv73} in the crust.

All numerical results presented here 
are obtained for a NS with the mass $M=1.4M_{\odot}$. 
The circumferential radius for such star is $R=12.2$~km,
the central density is 
$\rho_{\rm c}=9.26 \times 10^{14}$~g~cm$^{-3}$. 
The crust-core interface lies at the
distance $R_{\rm cc}=10.9$~km from the centre.

When modeling the effects of superfluidity we assume
the triplet pairing of neutrons 
and singlet  pairing of protons in the NS core.  
The neutron superfluidity in the stellar crust is neglected; 
it should not affect strongly
the global oscillations of NSs.

We consider two models of nucleon superfluidity: 
model `1'\  (simplified) and model `2'\ (more realistic). 
In the model 1 the redshifted
proton critical temperature 
is constant over the core, 
$T_{\rm cp}^\infty \equiv T_{\rm cp} \, {\rm e}^{\nu/2}=5\times 10^9$~K; 
the redshifted neutron critical temperature 
$T_{\rm cn}^\infty \equiv T_{\rm cn} \, {\rm e}^{\nu/2}$ 
increases with the density $\rho$ and reaches
the maximum value $T_{\rm cn\, max}^\infty=6\times 10^8$~K 
at the stellar centre ($r$=0). 
This model corresponds to the model 3 of \cite{kg11}.

In the model 2 both critical temperatures 
$T_{\rm cn}$ and $T_{\rm cp}$ 
are density dependent. 
This model does not contradict the results of
microscopic calculations  
(see, e.g., \citealt{ls01, yls99}) 
and is similar to the nucleon pairing models 
used to explain observations
of the cooling NS in Cassiopea A supernova remnant (\citealt{syhhp11}).

\begin{figure}
    \begin{center}
        \leavevmode
        \epsfxsize=6.4in \epsfbox{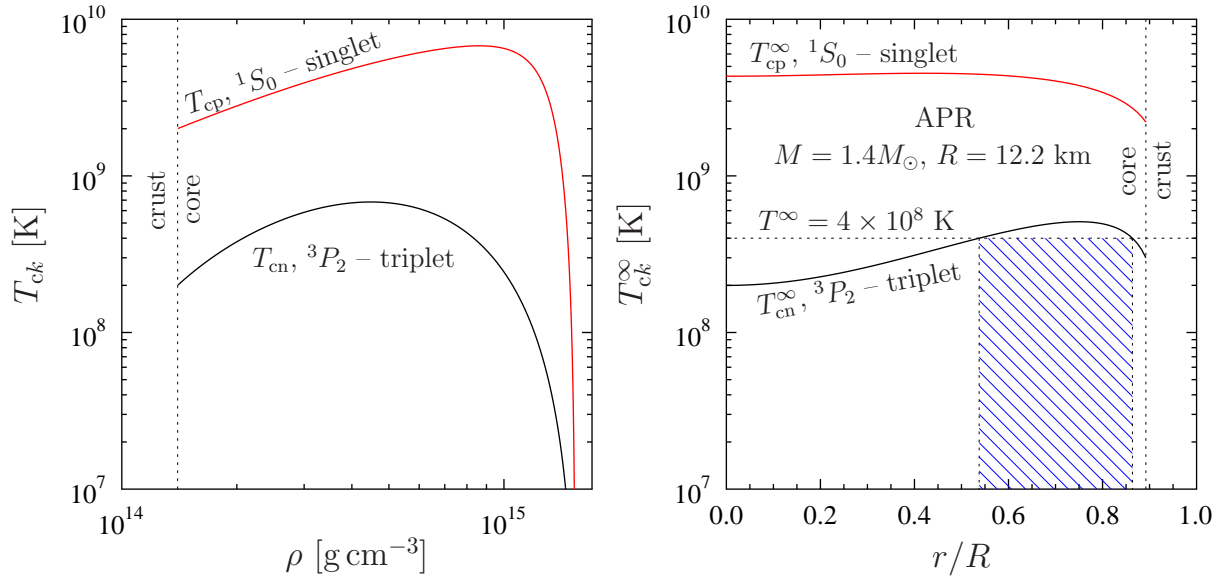}
    \end{center}
    \caption{(color online)
    The same as in Fig.\ \ref{Fig:Tc1} but for model 2.
    }
    \label{Fig:Tc2}
\end{figure}

The models 1 and 2 are shown 
in Figs.\ \ref{Fig:Tc1} and \ref{Fig:Tc2}, respectively
\footnote{Fig.\ \ref{Fig:Tc2} is a slightly modified version 
of Fig.\ 1 from \cite{cg11}.}.
The function $T_{{\rm c}i}(\rho)$ 
in both figures is shown in the left panels, 
while the right panels demonstrate 
the dependence $T_{{\rm c}i}^\infty(r)$ 
[$i={\rm n}$ and p]. 
With the decrease of the redshifted temperature $T^\infty$ 
the size of the SFL-region 
[given by the condition 
$T<T_{\rm cn}(r)$, or, equivalently, 
$T^\infty<T^\infty_{\rm cn}(r)$] 
increases or remains unchanged. 
For instance, 
the SFL-region corresponding to $T^\infty = 4\times10^8$~K, 
is shaded in Figs.\ \ref{Fig:Tc1} and \ref{Fig:Tc2}.
One can see that for the model 2 there can be three-layer configurations
of a star with no neutron superfluidity in the centre and in the
outer region but with superfluid intermediate region. 
On the contrary, in the model 1 
only two-layer configurations are possible.

The entrainment matrix $Y_{\rm ik}$
is calculated for the superfluidity models 1 and 2
in a way similar to how it was done in \cite{kg11}.

When analyzing viscous dissipation in oscillating NSs 
we allow for the damping due to shear and bulk viscosities.
For the shear viscosity coefficient $\eta$
we take the electron shear viscosity $\eta_{\rm e}$, 
calculated in \cite{sy08}.
We neglect the nucleon shear viscosity because:
($i$) it is poorly known even for nonsuperfluid matter and 
($ii$) it appears to be less than the electron shear viscosity
in the core at $T \ll T_{c {\rm p}}$ (\citealt{sy08}).

The bulk viscosity coefficients are calculated
as described by
\cite{gusakov07, gk08, kg11}. 
Since the direct URCA process is closed 
for our stellar model with $M=1.4 M_{\odot}$, 
the main contributor to the bulk viscosity is 
the modified URCA process.

\subsection{Oscillations of a nonsuperfluid star}
\label{SubSec:oscnormal}

As follows from Sec.\ \ref{SubSec:strategy}, 
before considering oscillations of a superfluid NS
one should study those of a normal
(nonsuperfluid) star of the same mass. 
%
To this aim, we have determined the eigenfrequencies and eigenfunctions of the
radial and nonradial oscillation modes
for a nonsuperfluid NS of mass $M=1.4\,M_\odot$ 
and equation of state APR (see Sec. \ref{SubSec:micro}). 
We have solved the equations 
describing radial and nonradial
perturbations of a nonrotating star in general relativity. 
These equations are derived by expanding 
the perturbed Einstein's equations 
in tensorial spherical harmonics 
in an appropriate gauge, 
and are integrated in the frequency domain.

Stellar modes are defined as solutions of the
perturbed equations which are regular at the centre 
and with vanishing Lagrangian pressure perturbation at the surface, 
and (if $l>1$) which behave as a pure outgoing wave at infinity;
as discussed above, such solutions have complex frequencies
$\omega=\sigma+{\rm i}/\tau$. 
If $l\le1$, instead, 
the frequency is real and the mode is
not associated to gravitational emission.

The oscillation modes 
are classified according to the source of the restoring force 
which prevails in bringing the perturbed element of fluid back
to the equilibrium position; 
for instance, we have a $g$-mode if the restoring
force is mainly provided by buoyancy, 
a $p$-mode if it is due to a gradient of pressure, 
and so on.

\begin{figure}
    \begin{center}
        \leavevmode
        \epsfxsize=6.4in \epsfbox{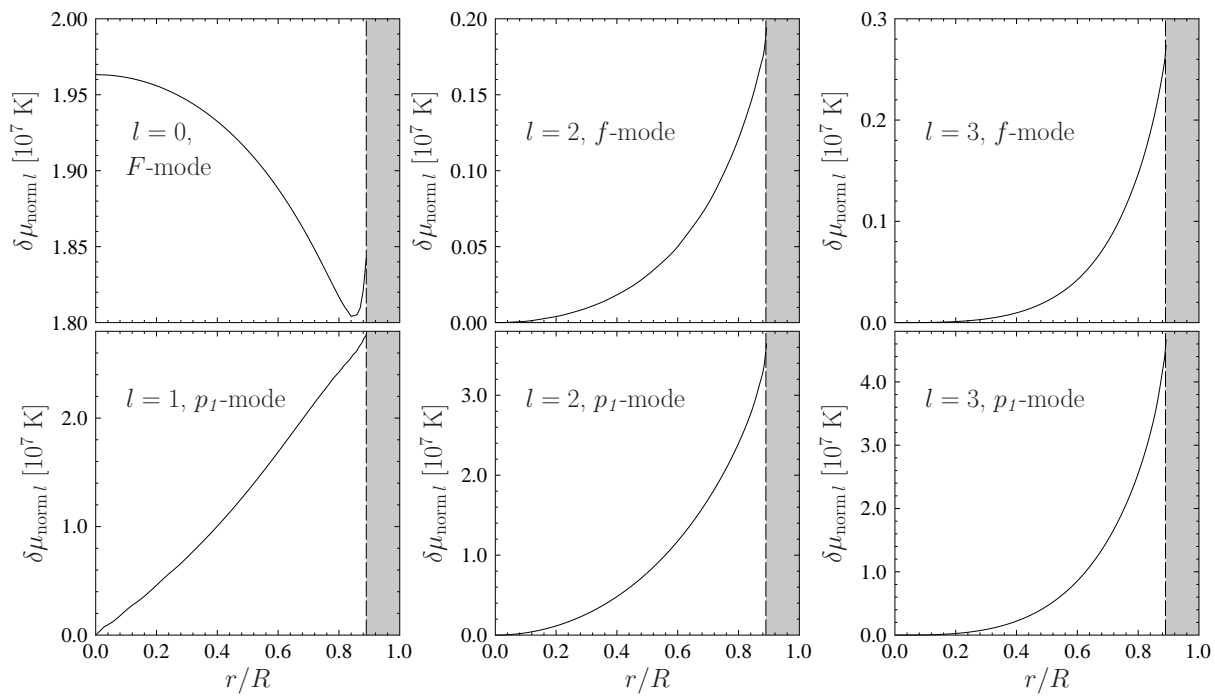}
    \end{center}
    \caption{
    The function $\delta \mu_{{\rm norm} \, l}$
    (in units of $10^7$ kelvins) versus $r$
    for fundamental radial $F$-mode as well as for
    $p_1$- and $f$-modes with multipolarities $l=1$, 2, and $3$
    (see the footnote $^{\ref{f_absent_footnote}}$).
    The energy of each oscillation mode is $10^{43}$ erg.
    Shaded region corresponds to crust,
    where $\delta \mu_{{\rm norm} \, l}$
    is not defined and was not plotted.
    }
    \label{Fig:dmu}
\end{figure}

\begin{figure}
    \begin{center}
        \leavevmode
        \epsfxsize=3.2in \epsfbox{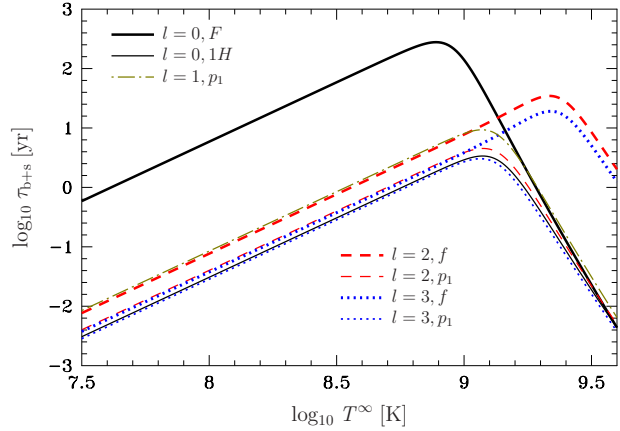}
    \end{center}
    \caption{(color online) Damping times
    $\tau_{\rm b+s} \equiv (\tau_{\rm bulk}^{-1}+\tau_{\rm shear}^{-1})^{-1}$
    versus $T^\infty$ for various oscillation modes.
    The effects of superfluidity are partially taken into account,
    as described in the text.
    Thick and thin solid lines correspond
    to radial ($l=0$) $F$- and $1\, H$-modes, respectively;
    dot-dashed line -- to dipole ($l=1$) $p_1$-mode;
    thick and thin dashes -- to quadrupole ($l=2$) $f$- and $p_1$-modes, respectively;
    thick and thin dots -- to octupole ($l=3$) $f$- and $p_1$-modes, respectively.
    }
    \label{Fig:tau_nsf}
\end{figure}

The radial modes are calculated as described in \cite{gyg05}.
To calculate the nonradial modes we follow the formulation 
of \cite{ld83} and \cite{dl85}.
In their formulation, 
the equations for nonradial perturbations can be expressed, 
inside the star, as a system of first-order 
differential equations in the variables 
$H_0, \,H_1,\, H_2,\, K,\, W_{\rm b}$, and $V_{\rm b}$ 
defined in Sec. \ref{SubSec:osc}. 
Outside the star, they reduce to a simple, 
second-order differential equation (the Zerilli equation). 
By numerical integration of these equations 
(the procedure we have followed is described in detail, e.g., in \citealt{bfgs11}) 
we find, for each value of 
the multipolarity $l$, 
the (complex) eigenfrequencies $\omega$ 
and the corresponding eigenfunctions 
$H_0(r)$, $H_1(r)$, $H_2(r)$, $K(r)$, $W_{\rm b}(r)$, and $V_{\rm b}(r)$. 
The results of our computations
are summarized in Table \ref{oscil} and illustrated
%
%
in Figs.\ \ref{Fig:dmu} and \ref{Fig:tau_nsf}.

\begin{table}
\caption{Frequency $\sigma$
(in units of $10^4$ s$^{-1}$ 
and in units of $\tilde{\sigma}=c/R \approx 2.46 \times 10^4$~s$^{-1}$)
and the damping time $\tau_{\rm grav}$ (in seconds)
for various oscillation modes of a nonsuperfluid NS.
The first column shows the multipolarity $l$
of modes and their names.}
\begin{center}
  \begin{tabular}{|l|c|c|c|}
  \hline \hline
        $l$, mode &  $\sigma/(10^4 \, {\rm s}^{-1})$ 
        & $\sigma/\tilde{\sigma}$ & $\tau_{\rm grav}$ (s) \\
  \hline
  0, $F$   & 1.703   & 0.691 & $\infty$    \\
  0, 1$H$  & 4.080   & 1.656 & $\infty$    \\
  0, 2$H$  & 5.732   & 2.327 & $\infty$    \\
  1, $p_1$ & 2.893   & 1.175 & $\infty$    \\
  2, $f$   & 1.155   & 0.469 & 0.212       \\
  2, $p_1$ & 3.720   & 1.510 & 3.799       \\
  3, $f$   & 1.554   & 0.631 & 18.24       \\
  3, $p_1$ & 4.360   & 1.770 & 33.26       \\
  \hline \hline
\end{tabular}
\label{oscil}
\end{center}
\end{table}

Table 1 presents the real parts of the eigenfrequencies 
${\rm Re}(\omega) = \sigma$ 
(measured in units of $10^4$ s$^{-1}$
and in units of 
$\tilde{\sigma} \equiv c/R \approx 2.46\times 10^4$ s$^{-1}$) 
and the characteristic gravitational 
damping times $\tau_{\rm grav}$ (in seconds) 
for the modes with 
$l=0$ (fundamental $F$-mode and
first two overtones $1\, H$ and $2\, H$), 
$l=1$ (dipole $p_1$-mode),
$l=2$ (quadrupole $f$- and $p_1$-modes), 
and $l=3$ (octupole $f$- and $p_1$-modes)
\footnote{The $f$-mode is absent in case of $l=1$. \label{f_absent_footnote}}.
One can see, that 
$\sigma \gg 1/\tau_{\rm grav}$ in all these cases.
That is, damping due to emission of gravitational waves
occurs on a time scale much longer than the oscillation period.

%

Using the definition (\ref{dmu5}) 
and Eq.\ (\ref{dmunorm_l}) we have determined, 
in terms of the eigenfunctions 
$H_0(r)$, $H_1(r)$, $\dots$, $V_{\rm b}(r)$, 
the function $\delta \mu_{{\rm norm}\, l}(r)$
and, consequently, the quantity
$\delta \mu^\infty_{\rm norm}(r,\, \theta)=
\delta \mu_{{\rm norm}\, l}(r) \, Y_l^0(\theta)$ 
for each mode. 
%
As follows from Eqs.\ (\ref{dmu3}) and (\ref{dmunorm}), 
for a {\it nonsuperfluid} star
$\delta \mu^\infty_{\rm norm}(r, \, \theta)$ 
is simply 
a redshifted imbalance of chemical potentials, 
$\delta \mu^\infty=\delta \mu_{\rm norm}^\infty$.
The function $\delta \mu_{{\rm norm}\, l}(r)$, 
entering Eq.\ (\ref{sfl3}),
is shown in Fig.\ \ref{Fig:dmu} 
for the oscillation modes from Table 1.
It is
normalized such that the mechanical energy 
of oscillations is $10^{43}$ erg. 
The shaded region corresponds to the crust of the star, 
where $\delta \mu_{{\rm norm} \, l}(r)$ is not defined 
(protons are bound in nuclei there). 
As seen in the figure, 
$|\delta \mu_{{\rm norm} \, l}(r)|$ for $f$-modes
is about one order of magnitude smaller than for $p$-modes. 
This is not surprising, 
since matter is only weakly compressed 
during $f$-mode oscillations, 
so that a deviation from beta-equilibrium 
(when $\delta \mu^\infty=0$) 
is small. 
The functions $\delta \mu_{{\rm norm} \,l}(r)$ 
are employed to calculate the damping times 
of a superfluid NS
in Sec.\ \ref{SubSec:damping}.

Fig.\ \ref{Fig:tau_nsf} shows the 
viscous damping time 
$\tau_{\rm b+s} \equiv (\tau_{\rm bulk}^{-1}+\tau_{\rm shear}^{-1})^{-1}$ 
as a function of $T^\infty$ for a set of oscillation modes.
The solid lines corresponds 
to radial  ($l=0$) modes $F$ and $1\, H$;
dot-dashed line to dipole ($l=1$) mode $p_1$;
dashed lines to quadrupole  ($l=2$) modes $f$ and $p_1$;
dotted lines to octupole ($l=3$) modes $f$ and $p_1$.
To calculate $\tau_{\rm b+s}$ 
we used the formulas 
for $\tau_{\rm bulk}$ and $\tau_{\rm shear}$, 
applicable for the ordinary hydrodynamics 
of a nonsuperfluid liquid
\footnote{More precisely, 
we used Eqs.\ (\ref{taubulk2}) and (\ref{taushear2})
with $W_{\rm sfl}=V_{\rm sfl}=0$.}.
However, we allow for the effects of superfluidity when calculating 
the kinetic coefficients $\eta$ and $\xi_2$
(the other bulk viscous coefficients
do not appear in the normal fluid hydrodynamics).
To calculate  $\eta$ and $\xi_2$ 
we adopt the nucleon superfluidity model 2
(see Sec.\ \ref{SubSec:micro}). 
Such an approximate approach 
to accounting for the effects of superfluidity
is commonly used in the literature, 
but it is not fully consistent.
The results of a more consistent approach 
(see Sec.\ \ref{Sec:approach}) 
are discussed below in Sec.\ \ref{SubSec:damping}.

As follows from Fig.\ \ref{Fig:tau_nsf},
the dependence of $\tau_{\rm b+s}$ on $T^\infty$ 
is a power-law at $T^\infty \la 6 \times 10^8$~K.
At such $T^\infty$ the proton superfluidity is `strong'\
($T^\infty \ll T_{\rm cp}^\infty$).
In that case the bulk viscosity 
is exponentially suppressed (\citealt*{hly01}), 
while the shear viscosity
$\eta \propto 1/(T^\infty)^2$ (\citealt{sy08}) and dominates.
As a result, 
$\tau_{\rm b+s} \propto (T^\infty)^2$.
At high enough temperatures 
$T^\infty \ga 6 \times 10^8$~K 
the damping due to the bulk viscosity starts to prevail;
this results in decreasing of $\tau_{\rm b+s}$ with growing $T^{\infty}$ 
(the curves in Fig.\ \ref{Fig:tau_nsf} bend down).
At such $T^\infty$ 
the neutrons are normal 
and the proton superfluidity is weak or absent.
Neglecting the proton superfluidity, 
one obtains $\xi_2 \propto (T^\infty)^6$ (\citealt{hly01}), 
hence $\tau_{\rm b+s} \propto 1/(T^\infty)^6$.

Let us note that the curves for $f$-modes
in Fig.\ \ref{Fig:tau_nsf} 
(thick dashed line and thick dots) 
bend down later than others;
for them the shear viscosity is the dominant mechanism
of damping up to $T^\infty \approx 2.0 \times 10^9$~K. 
This is not surprising, 
since, as it was noted above, 
for $f$-modes the deviation from beta-equilibrium is small 
($\delta \mu^\infty$ is reduced by an order of magnitude 
in comparison to $p$-modes, 
see Fig.\ \ref{Fig:dmu}),
hence damping due to the bulk viscosity 
is suppressed
(the relation between 
$\delta \mu^\infty$ and $\tau_{\rm bulk}$
was discussed in detail, 
e.g., in \citealt{gyg05}).
As a result, 
$\tau_{\rm b+s}$ approaches its `bulk viscosity' asymptote
$\tau_{\rm b+s} \propto 1/(T^\infty)^6$
at higher temperatures $T^\infty>2.0 \times 10^9$~K.


\subsection{Frequency spectrum for superfluid NSs}
\label{SubSec:spectra}

First of all let us consider
the frequency spectrum for radial oscillations 
of a superfluid neutron star employing
the simplified model 1 of nucleon superfluidity.
For such model this problem 
was discussed in detail by \cite{kg11}, 
where it was solved exactly.
Here we compare this exact solution 
with the approximate calculations 
obtained in the $s=0$ approximation 
(see Sec.\ \ref{Sec:approach}).
Such a comparison is very useful, 
since it allows one 
to make a conclusion about applicability 
of the approximate approach 
in the case of nonradial oscillations, 
where the exact solution is not attempted.

\begin{figure}
    \begin{center}
        \leavevmode
        \epsfxsize=6.4in \epsfbox{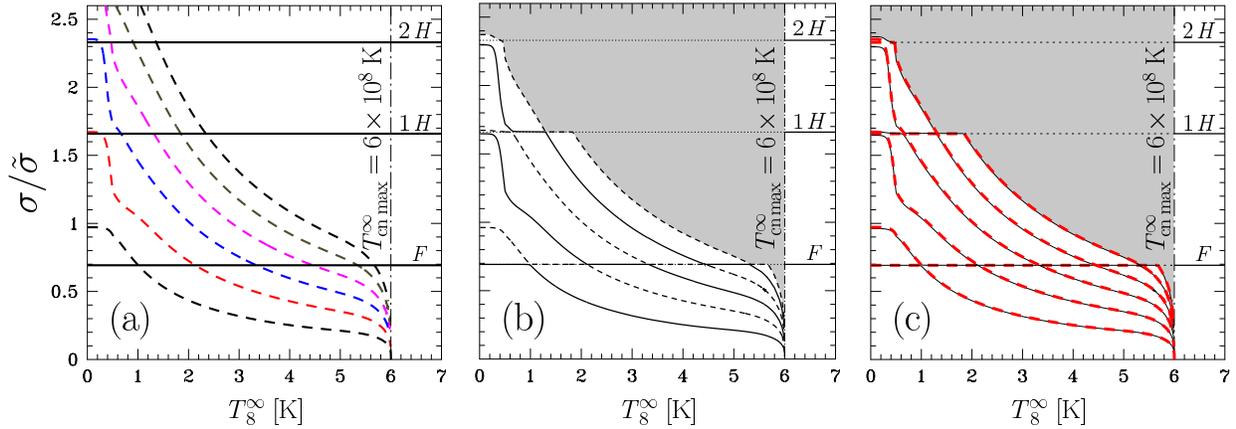}
    \end{center}
    \caption{(color online) Eigenfrequencies $\sigma$
    (in units of $\tilde{\sigma}$) of radial oscillations
    versus $T^\infty_8=T^\infty/(10^8 \, {\rm K})$
    for model 1 of nucleon superfluidity.
    (a) Approximate spectrum.
    First three normal modes ($F$, $1\, H$, and $2\, H$) are shown by the solid lines;
    first six superfluid modes $1,\ldots,6$ are shown by dashes.
    (b) Exact spectrum.
    Alternate solid and dashed lines show the first six
    exact modes (${\rm I,\ldots,VI}$) of a radially oscillating star.
    No spectrum was plotted in the shaded region.
    (c) Approximate (dashed lines) and exact (solid lines) spectra.
    At $T^\infty>T_{\rm cn \, max}^\infty=6 \times 10^8$~K all neutrons are normal
    and the spectrum is that of a nonsuperfluid star.
    }
    \label{Fig:w_rad}
\end{figure}

The eigenfrequencies $\sigma$ of radial pulsations
(in units of $\tilde{\sigma}$) 
versus $T^\infty_8=T^\infty/(10^8$~K) 
are shown in Fig.\  \ref{Fig:w_rad}(a, b, c). 
In Fig.\ \ref{Fig:w_rad}(a) 
this dependence was obtained assuming 
that superfluid and normal modes are completely decoupled 
($s=0$ approximation).
The thick solid lines demonstrate 
the first three normal (nonsuperfluid) radial modes 
$F$, $1 \,H$, and $2 \,H$.
As one expects, 
their frequencies do not depend on $T^\infty$.
The dashes are for the first six superfluid modes $1,\ldots,6$, 
which are the solutions to Eq.\ (\ref{sfl4}).
These modes, on the contrary, 
strongly depend on $T^\infty$ 
and approach their temperature-independent asymptotes 
only at $T^\infty \la 5 \times 10^7$~K
(when the entire NS core is superfluid 
and $Y_{ik}$ does not depend on $T^\infty$).
At $T^\infty >T_{\rm cn \, max}^{\infty}=6 \times 10^8$~K
all neutrons are normal so that superfluid modes do not exist.

%
%

Fig.\ \ref{Fig:w_rad}(b) demonstrates 
the results of the exact solution 
to Eqs.\ (\ref{dnb})--(\ref{dmunorm_l}) 
obtained by \cite {kg11} 
for radial oscillations of a superfluid neutron star. 
The frequencies $\sigma$ 
of the first six oscillation modes 
(I,$\ldots$,VI)
as functions of $T^\infty_8$ 
are shown by alternate solid and dashed lines.
No spectrum is plotted 
in the gray-shaded area.
One can observe that 
the approximate spectrum [Fig.\ \ref{Fig:w_rad}(a)]
is very similar to the exact spectrum [Fig.\ \ref{Fig:w_rad}(b)].
However, there is one important difference:
instead of \textit{crossings} of superfluid and normal
modes in Fig.\ \ref{Fig:w_rad}(a) 
we have \textit{avoided crossings} 
of the modes in Fig.\ \ref{Fig:w_rad}(b).
At these points the superfluid mode 
turns into the normal one and vice versa. 
As it was discussed 
in details in \cite{gk11, kg11},
this is not surprising, 
since in a vicinity of avoided crossings 
the Einstein equations (\ref{einst})
and superfluid equation (\ref{sfl3}) 
interact resonantly,  
so that approximation 
of completely decoupled superfluid and normal modes ($s=0$) 
is inapplicable
\footnote{Thus, it would not be correct to say 
that any \textit{real} oscillation mode of a superfluid star 
is either purely superfuid or purely normal:
for some $T^\infty$ 
it can 
show itself as
a superfluid, 
but for other $T^\infty$ it can behave as a normal mode
[see Fig.\ \ref{Fig:w_rad}(b)].}.

For comparison, 
in Fig.\ \ref{Fig:w_rad}(c) we plot
both the approximate (dashed lines)
and exact (solid lines) spectra. 
The agreement between both spectra is very good: 
the difference is less than a few per cent.

\begin{figure}
    \begin{center}
        \leavevmode
        \epsfxsize=5in \epsfbox{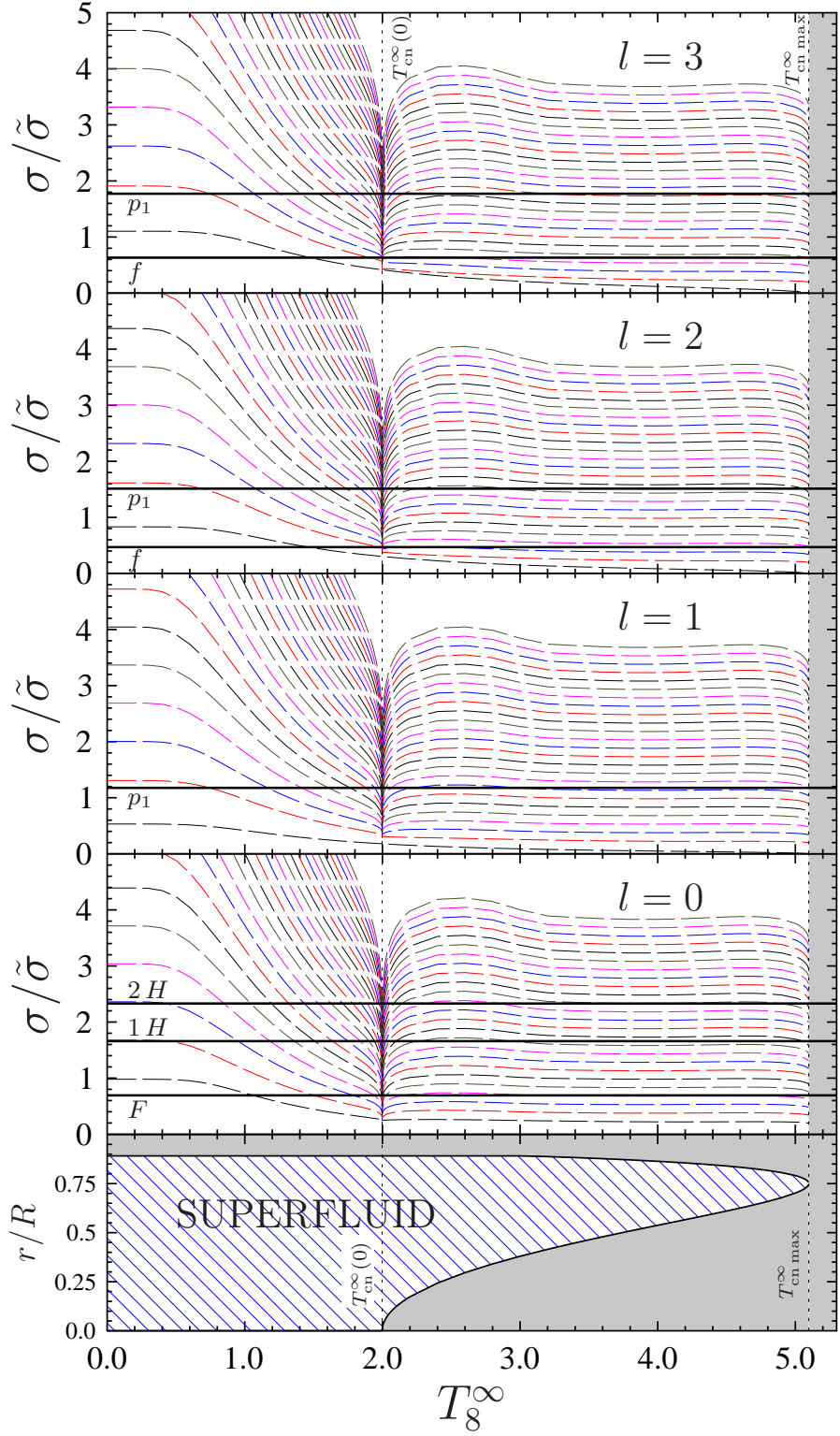}
    \end{center}
    \caption{(color online) Eigenfrequencies $\sigma$ versus $T^\infty$
    for model 2 of nucleon superfluidity and for multipolarities $l=0$, $1$, $2$, and $3$.
    For each $l$ we plot first few normal modes (solid lines)
    and first 25 superfluid modes (dashed lines), 
    whose eigenfunctions $\delta \mu_l$ differ by the number of radial nodes $n$.
    At $T^\infty \leq T^\infty_{\rm cn}(0) \approx 2 \times 10^8$~K
    (see the left vertical dotted line),
    neutron superfluidity occupies the stellar centre.
    The bottom panel demonstrates the variation of the SFL-region (shown by hatching) with $T^\infty$.
    Shaded area in all panels shows the region where all neutrons are normal.
    }
    \label{Fig:w_nonrad}
\end{figure}

Such a close agreement 
of the exact and approximate results 
for radial oscillations 
allows us to analyse the spectrum of nonradial oscillations 
using the same approximation $s=0$. 
The results of this analysis 
are shown in Fig.\ \ref{Fig:w_nonrad} 
for more realistic model 2 of nucleon superfluidity 
(see Sec.\ \ref{SubSec:micro} and Fig.\ \ref{Fig:Tc2}). 
Superfluid modes 
shown in this figure
have been already studied
in detail
in our recent paper (\citealt{cg11}).
Thus, here we discuss them only briefly.

Fig.\ \ref{Fig:w_nonrad} contains five panels. 
Four upper panels present eigenfrequencies $\sigma$ 
as functions of $T^\infty_8$ 
for normal modes from Table 1 
(thick horizontal lines)
and for superfluid modes (dashes) 
with multipolarities $l=0$, 1, 2, and 3.
For each $l$ there is an infinite set of superfluid modes 
whose eigenfunctions $\delta \mu_l$ differ by the number of radial nodes $n$;
in the figure we plot the first 25 of them.
The lower panel 
demonstrates broadening 
of the SFL-region
with decreasing $T^\infty_8$ 
(SFL-region is shown by hatches).
For model 2 (which we employ here) 
the redshifted neutron critical temperature $T_{\rm cn}^\infty(r)$
has a maximum at 
$T_{\rm cn \, max}^\infty \approx 5.1 \times 10^8$~K
(right vertical dotted line).
The neutron superfluidity reaches the stellar centre 
at $T^\infty = T_{\rm cn}^\infty(0) \approx 2 \times 10^{\rm 8}$~K
(left vertical dotted line).
At $T^\infty>T_{\rm cn \, max}^\infty$ 
all neutrons are normal, 
hence only normal modes exist in the star. 
At $T^\infty < T_{\rm cn}^\infty(0)$ 
the core is completely occupied by the neutron superfluidity.
One can see that the behaviour of superfluid modes 
differs strongly at $T^\infty>T_{\rm cn}^\infty(0)$ and
at $T^\infty<T_{\rm cn}^\infty(0)$. 
This feature was discussed in \cite{cg11, kg11}, 
where it was demonstrated that (roughly speaking) 
the frequencies $\sigma$ of superfluid modes
scale with $Y_{\rm nn}$ and $R_{\rm sfl}$ as
$\sigma \sim \sqrt{Y_{\rm nn}}/R_{\rm sfl}$, 
where $R_{\rm sfl}$ is the size of the SFL-region.
With the increasing of temperature $Y_{\rm nn}$ decreases, 
while the size of the SFL-region 
can either decrease 
[at $T^\infty>T_{\rm cn}^\infty(0)$] 
or remain constant [at $T^\infty<T_{\rm cn}^\infty(0)$].
As a result, 
there is a partial compensation 
of these two tendencies at $T^\infty>T_{\rm cn}^\infty(0)$ 
(hence, the frequency changes only weakly),
while at $T^\infty<T_{\rm cn}^\infty(0)$ 
the effect of decreasing of $Y_{\rm nn}$
is not compensated 
(hence, $\sigma$ decreases with growing $T^\infty$).
At $T^\infty \la 5 \times 10^{7}$~K
$Y_{ik}$ does not depend on $T^\infty$,
and, as in the case of radial pulsations,
the frequencies approach 
their low-temperature asymptotes.

\subsection{Damping times for superfluid NSs}
\label{SubSec:damping}

As in the case of eigenfrequencies,
we first consider the e-folding times
$\tau_{\rm b+s}^{-1} \equiv \tau_{\rm bulk}^{-1}+\tau_{\rm shear}^{-1}$
for radial ($l=0$) pulsations for the simplified model 1
of nucleon superfluidity (see Fig.\ \ref{Fig:Tc1}).

\begin{figure}
    \begin{center}
        \leavevmode
        \epsfxsize=6.4in \epsfbox{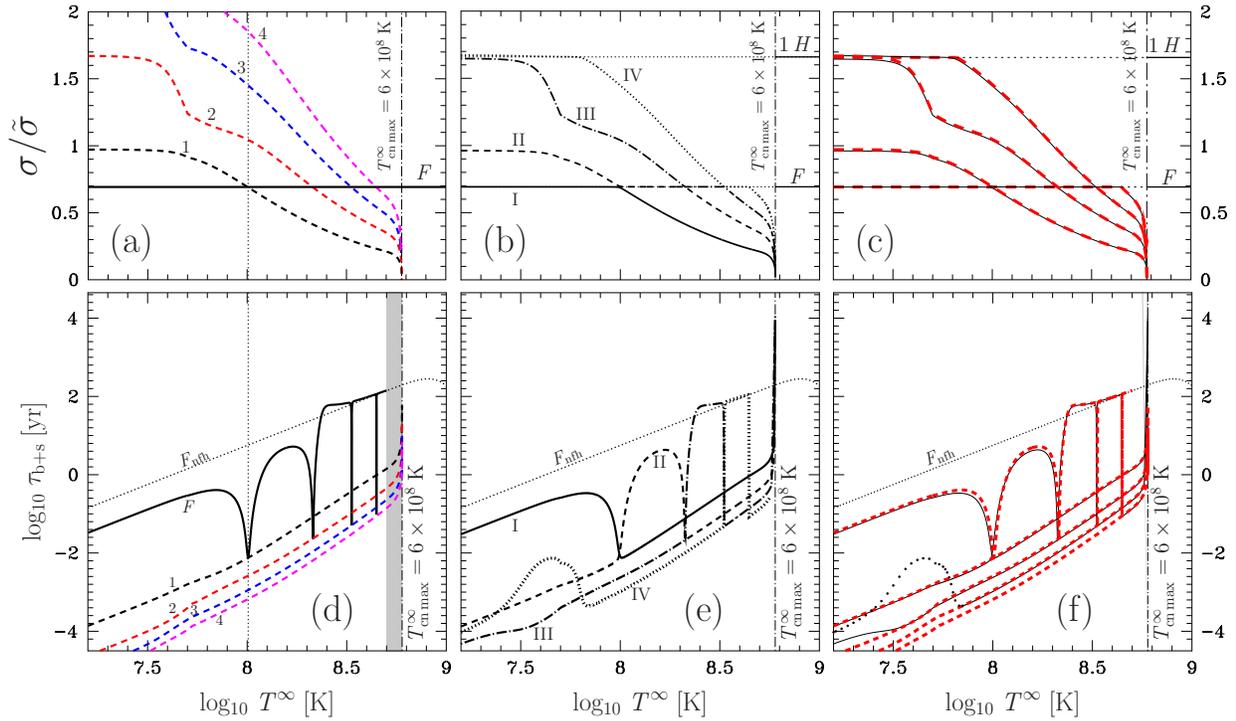}
    \end{center}
    \caption{(color online)
    Eigenfrequencies $\sigma$ [panels (a, b, c)] 
    and damping times $\tau_{\rm b+s}$ [panels (d, e, f)] 
    of a radially oscillating NS versus $T^\infty$ 
    for model 1 of nucleon superfluidity. 
    Panels (a, d): Approximate solution
    (normal $F$-mode and first four superfluid modes $1,\ldots,4$
    are shown by solid and dashed lines, respectively);
    Panels (b, e): Exact solution (first four exact modes ${\rm I},\ldots,{\rm IV}$ 
     are shown by solid, dashed, dot-dashed, and dotted lines, respectively); 
    Panels (c, f): Both approximate (dashed lines) and exact (solid lines) solutions. 
    Panels (a, b, c) are the same spectra 
    as those plotted, respectively, in Fig.\ \ref{Fig:w_rad}(a, b, c).
    Normal $F$-mode is not shown in the shaded region because of technical reasons 
    (too many resonances).
    Dotted lines in panels (d, e, f) show damping times for $F$-mode calculated
    using ordinary normal-fluid hydrodynamics (see the text for more details).
    Part of the mode IV (at $T^\infty <6 \times 10^7$~K) 
    is shown by dots in the panel (f), as described in the text.
    }
    \label{Fig:tau_rad}
\end{figure}

In Fig.\ \ref{Fig:tau_rad}(a, d) we present 
the functions $\sigma(T^\infty)$ and $\tau_{\rm b+s}(T^\infty)$,
obtained using the approximate method 
of Sec.\ \ref{SubSec:strategy}.
The frequencies and damping times are plotted
for normal $F$-mode (thick solid line)
as well as for the first four superfluid modes 
$1,\ldots,4$ (dashed lines)
\footnote{Notice that, in Figs.\ \ref{Fig:tau_rad}(a, b, c)
we present, in logarithmic scale, 
parts of the spectra, 
which were already plotted in linear scale 
in Figs.\ \ref{Fig:w_rad}(a, b, c), respectively.}.
In the region shaded in gray the function 
$\tau_{\rm b+s}(T^\infty)$ 
for the normal mode was not plotted 
(there are too many merging resonances in this region).
The dotted curve in Fig.\ \ref{Fig:tau_rad}(d, e, f)
labeled $F_{\rm nfh}$ 
(`nfh' is the abbreviation for `normal-fluid hydrodynamics')
shows the damping time
calculated using the ordinary hydrodynamics of nonsuperfluid liquid
but taking into account the effects of superfluidity
on the bulk and shear viscosities.
This curve is analogous to 
the thick solid curve in Fig.\ \ref{Fig:tau_nsf},
obtained under the same conditions but for the model 2 
of nucleon superfluidity.
The vertical dotted line in Fig.\ \ref{Fig:tau_rad}(a, d)
indicates a temperature at which 
frequencies of normal $F$-mode 
and the first superfluid mode coincide.

\begin{figure}
    \begin{center}
        \leavevmode
        \epsfxsize=6.4in \epsfbox{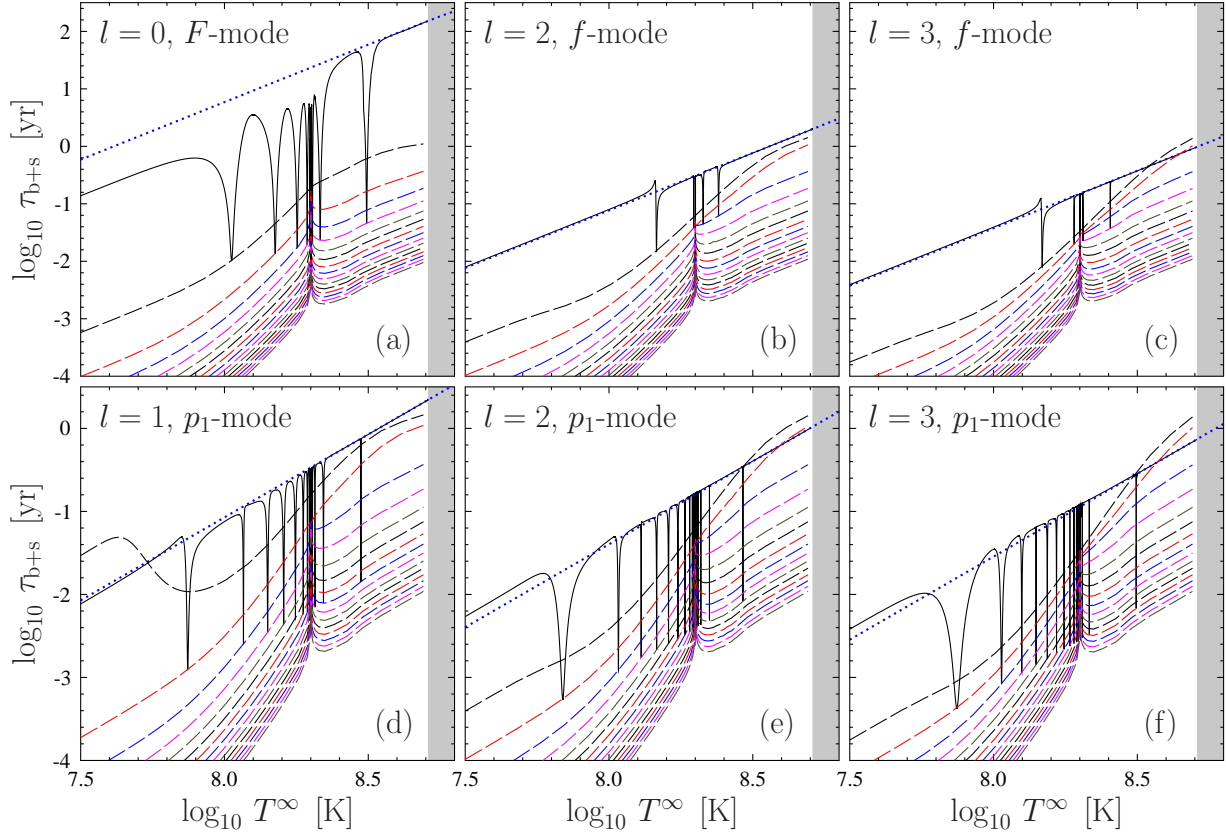}
    \end{center}
    \caption{(color online) Damping times $\tau_{\rm b+s}$ versus $T^\infty$ 
    for various oscillation modes for model 2 of nucleon superfluidity.
    On each panel we plot one normal mode 
    (shown by solid line; its multipolarity and name is indicated)
    and first 15 superfluid modes (dashed lines).
    Dotted lines show $\tau_{\rm b+s}(T^\infty)$ for normal modes calculated
    using normal-fluid hydrodynamics 
    [see the text 
    for more details].
    In the shaded area all neutrons are normal and superfluid modes do not exist.
    }
    \label{Fig:tau_nonrad}
\end{figure}

\begin{figure}
    \begin{center}
        \leavevmode
        \epsfxsize=6.4in \epsfbox{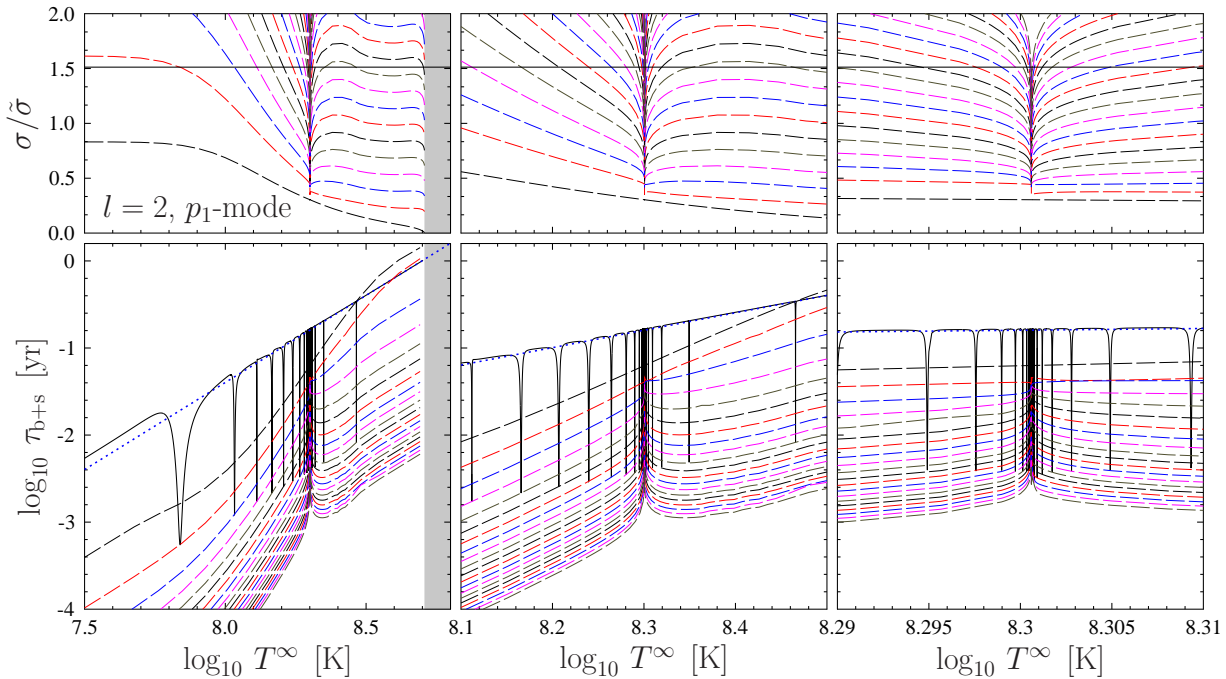}
    \end{center}
    \caption{(color online) 
    Eigenfrequencies $\sigma$ (upper panels) and damping times $\tau_{\rm b+s}$ (lower panels)
    versus $T^\infty$ for quadrupole ($l=2$) oscillation modes 
    in an increasingly larger scale. 
    The normal $p_1$-mode is shown by solid lines.
    Left lower panel coincides with Fig.\ \ref{Fig:tau_nonrad}(e). 
    Other lower panels are zoomed in versions 
    of Fig.\ \ref{Fig:tau_nonrad}(e).
    Notations are the same as 
    in Figs.\ \ref{Fig:w_nonrad} and \ref{Fig:tau_nonrad}.
    }
    \label{Fig:tau_l2p1}
\end{figure}

We present a detailed analysis of Fig.\ \ref{Fig:tau_rad}(d)
in what follows, 
together with description of the approximate 
solutions for nonradial oscillation modes 
(Figs.\ \ref{Fig:tau_nonrad} and \ref{Fig:tau_l2p1}).

For comparison, 
Fig.\ \ref{Fig:tau_rad}(b, e) 
demonstrates the results of the exact calculation 
of frequencies $\sigma(T^\infty)$
and damping times $\tau_{\rm b+s}(T^\infty)$ 
for the first four (I,$\ldots$,IV)
oscillation modes of the superfluid NS
[the modes are shown by solid (I), 
dashed (II), dot-dashed (III), and dotted (IV) lines].

To see how well the approximate solution 
[Fig.\ \ref{Fig:tau_rad}(a, d)]
agrees with the exact one
[Fig.\ \ref{Fig:tau_rad}(b, e)], 
both solutions are presented in Fig.\ \ref{Fig:tau_rad}(c, f).
Dashes correspond to approximate solution,
solid lines -- to exact solution.
A portion of the mode IV in Fig.\ \ref{Fig:tau_rad}(f)
is shown by thick dots because 
the corresponding approximate solution (the mode $1 \, H$)
is not plotted.
One sees that the agreement between the approximate 
and exact solutions is reasonable 
everywhere (average error does not exceed $10-25\%$) 
except for the resonances (see below) and 
an interval of temperatures $T^\infty \la 3\times 10^7$~K
where the mode III of exact solution deviates from 
the second superfluid mode of approximate solution.
To explain this deviation let us note that,
as follows from Fig.\ \ref{Fig:w_rad}(a), 
at such $T^\infty$ the frequency of the 
normal mode $1 \,H$ practically coincides
with that of the second superfluid mode.
In that case Eqs.\ (\ref{einst}) and (\ref{sfl3}) 
interact resonantly, 
so that the approximation
of independent superfluid and normal modes is poor
even though parameter $s$ is small
%
\footnote{For a quantitatively correct description
of the function $\tau_{\rm b+s}(T^\infty)$
in that case it is, in principle, 
straightforward to develop a perturbation theory
similar to the degenerate perturbation theory of quantum mechanics
(see below the discussion of resonances 
in Figs.\ \ref{Fig:tau_rad}, \ref{Fig:tau_nonrad}, and \ref{Fig:tau_l2p1}).}.
%

Let us now consider the nonradial oscillations.
Fig.\ \ref{Fig:tau_nonrad} presents an approximate solution
for the function $\tau_{\rm b+s}(T^\infty)$,
which is obtained for a realistic nucleon superfluidity model 2.
By dashes we show superfluid modes, 
solid lines correspond to normal modes.
Each panel in the figure is plotted for one normal mode
(its name and multipolarity $l$ are indicated)
and for the first 15 superfluid modes with the same $l$.
By dots, as in Fig.\ \ref{Fig:tau_rad}(d, e, f), 
we plot $\tau_{\rm b+s}$ for a corresponding normal modes 
calculated using the ordinary normal-fluid hydrodynamics.
In the shaded region superfluid modes were not plotted
because all neutrons are normal there
and the star oscillates as a nonsuperfluid.

In more detail damping times 
are demonstrated for quadrupole ($l=2$) 
oscillation modes in Fig.\ \ref{Fig:tau_l2p1}.
In particular, the normal $p_1$-mode is 
shown there by solid lines.
In the three lower panels we plot 
the dependence $\tau_{\rm b+s}(T^\infty)$
in an increasingly larger scale.
In the three upper panels 
we plot, in the same scale, 
the oscillation frequencies $\sigma(T^\infty)$ 
(the corresponding spectrum was already presented 
in Fig.\ \ref{Fig:w_nonrad} in linear scale).
Left lower panel of Fig.\ \ref{Fig:tau_l2p1} 
coincides with Fig.\ \ref{Fig:tau_nonrad}(e).

Let us discuss the main conclusions that can be drawn 
from the analysis of 
Figs.\ \ref{Fig:tau_rad}(d), \ref{Fig:tau_nonrad}, and \ref{Fig:tau_l2p1}.

1. For any normal mode 
the dependence $\tau_{\rm b+s}(T^\infty)$
(solid lines in these figures),
has a set of {\it resonance features} (spikes) concentrated
(for radial and $p$-modes)
to the critical temperature 
$T_{\rm cn}^\infty(0)$ 
at which neutron superfluidity 
in the core centre dies out.
 For model 1 $T_{\rm cn}^\infty(0)
 =T_{\rm cn \, max}^\infty=6 \times 10^8$~K (see Fig.\ \ref{Fig:Tc1}),
for model 2
$T_{\rm cn}^\infty(0) \approx 2 \times 10^8$~K 
(see Fig.\ \ref{Fig:Tc2}).
The resonances appear when frequency of the normal mode
approaches the frequency of one of the superfluid modes.
For instance, solid line in Fig.\ \ref{Fig:tau_rad}(a)
crosses superfluid modes four times
[in Fig.\ \ref{Fig:tau_rad}(a, d) 
the temperature $T^\infty$ of the first crossing 
is shown by the vertical dotted line 
and equals $T^\infty \approx 10^8$~K].
Correspondingly, four resonances appear in Fig.\ \ref{Fig:tau_rad}(d).
A similar situation can be observed 
in Figs.\ \ref{Fig:tau_nonrad} and \ref{Fig:tau_l2p1}.
Near resonances $\tau_{\rm b+s}$ for normal mode 
rapidly decreases by 1--2 orders of magnitude (see item 2 below) and, 
in the resonance point, 
it becomes strictly equal to
$\tau_{\rm b+s}$ for the corresponding superfluid mode.

Such behavior of the approximate solution
$\tau_{\rm b+s}(T^\infty)$ for normal modes 
in the vicinity of resonances can be easily understood.
In resonance points, in which the frequencies of superfluid and normal modes coincide,
Eq.\ (\ref{sfl3}) has a nontrivial solution even 
in the absence of the source $\delta \mu_{{\rm norm} \, l}$. 
For it to be satisfied with the source, 
the oscillation amplitude $\delta \mu$
must be infinitely large.
In other words,
in resonance points all the energy must be contained 
in superfluid degrees of freedom
(in particular, near resonances 
$W_{\rm sfl} \gg W_{\rm b}$ and $V_{\rm sfl} \gg V_{\rm b}$).
Formally, this means that in the resonance point 
the damping time $\tau_{\rm b+s}$
should be exactly the same as for the superfluid mode.

Another important point that is worth noting is that,
as follows from Fig.\ \ref{Fig:tau_rad}(f), 
the approximate solution for the normal radial $F$-mode 
describes {\it qualitatively well}
the exact solution {\it near resonances} 
(the latter is shown by solid lines).
We expect that the same is also true for nonradial modes 
for which the exact solution was not attempted.
At first glance 
such 
an agreement between 
the approximate and exact solutions
seems surprising
because the approximation $s=0$ should not work
in the vicinity of resonances, 
where the frequencies of superfluid and normal modes 
are close to each other.
Nevertheless, one verifies that 
this approximation 
is still suitable for a qualitatively correct
description of the function $\tau_{\rm b+s}(T^\infty)$
if one bears in mind that:
($i$) close to any resonance the exact solution
is a linear superposition of independent solutions
describing (intersecting) superfluid and normal modes and
($ii$) $\tau_{\rm b+s}$ for the superfluid mode
is much less than for the normal mode.

Items ($i$) and ($ii$) mean that,
in the exact solution, 
the main contribution to $\tau_{\rm b+s}$
comes from the superfluid mode 
(while the contribution from the normal mode is small).
This leads us to conclusion that the 
superfluid modes are the main sources of viscous dissipation 
in the vicinity of resonance points.
The same conclusion was already drawn above 
using the approximate method of Sec.\ \ref{SubSec:strategy}.
%
This explains why 
the approximate method 
gives qualitatively correct results for 
$\tau_{\rm b+s}(T^\infty)$ 
near resonances.

In order to avoid confusion
let us emphasize that the function $\tau_{\rm b+s}(T^\infty)$ 
contains resonance features (spikes) 
for normal modes 
only in the approximate solution 
[see Figs.\ \ref{Fig:tau_rad}(d), \ref{Fig:tau_nonrad}, and \ref{Fig:tau_l2p1}].
In the exact solution any normal oscillation mode 
turns into a superfluid one near resonance (and vice versa).
This leads to an abrupt decreasing (increasing) of $\tau_{\rm b+s}$
and formation of a `step-like' structure 
rather than spike [see Fig.\ \ref{Fig:tau_rad}(e)].

2. It was already mentioned above that, 
as follows from 
Figs.\ \ref{Fig:tau_rad}(d), \ref{Fig:tau_nonrad}, and \ref{Fig:tau_l2p1}, 
normal modes (far from resonances) damp out by 1--2 orders of magnitude slower
than those superfluid modes with which they can have equal frequencies 
(i.e. intersect in the $\sigma-T^{\infty}$ plane).

What is the reason for such a fast damping of superfluid modes?
To be more concrete, below we consider 
a low-temperature case, 
$T^\infty \la 3 \times 10^7$~K.
There are three main factors:
($i$) For superfluid modes eigenfunctions $W_{\rm sfl}$ and $V_{\rm sfl}$
have a maximum in the central regions of a star 
where the shear viscosity is maximal.
On the contrary, for normal modes 
the maximum of eigenfunctions $W_{\rm b}$ and $V_{\rm b}$
lies closer to the NS surface, 
where the shear viscosity coefficient
can be substantially (5 and more times) smaller.
As a consequence, $\mathfrak{W}_{\rm shear}$
for superfluid modes turns out to be greater
(and hence $\tau_{\rm shear}$ smaller) 
than for normal modes.
($ii$) The energy of superfluid modes is given by Eq.\ (\ref{Emechsfl})
and depends on the quantity $y$ 
[see Eq.\ (\ref{y}) for the definition of $y$].
At low $T^\infty$ the parameter $y$ is small,
$y \sim n_{\rm p}/n_{\rm n} \sim 0.04 \div 0.09$, 
which also results in decreasing of the 
characteristic damping times for superfluid modes
\footnote{
To get an estimate for $y$ we made use of the sum rule
$\mu_{\rm n}Y_{\rm nn}+\mu_{\rm p} Y_{\rm np}=n_{\rm n}$
valid at $T^\infty=0$  (\citealt*{gkh09a}),
and neglected the small matrix element $Y_{\rm np}$
in comparison with $Y_{\rm nn}$.
}.
($iii$) This factor is particularly important for radial oscillations ($l=0$)
and is related to a coefficient $\alpha_1$ in the expression (\ref{taushear2}) 
for the damping time $\tau_{\rm shear}$ due to shear viscosity. 
This coefficient is given by Eq.\ (\ref{alpha1})
which is a sum of four terms. 
It turns out that for the normal radial modes 
the first term is well compensated by the third term $H_2$, 
while the other therms vanish.
For the superfluid modes such compensation does not occur
because for them $H_2=0$.

3. At low enough $T^\infty$
the damping times 
for normal radial and $p$-modes 
can be several times 
larger or smaller that 
$\tau_{\rm b+s}$, 
calculated employing ordinary hydrodynamics of nonsuperfluid liquid
but accounting for the effects of superfluidity on the 
bulk and shear viscosity coefficients
(dotted lines in Figs.\ \ref{Fig:tau_rad}(d), 
\ref{Fig:tau_nonrad}(a, d, e, f), and \ref{Fig:tau_l2p1}).
Let us inspect, for example, Fig.\ \ref{Fig:tau_nonrad}(a). 
One sees that at $T^\infty \la 10^8$~K 
$\tau_{\rm b+s}$, 
calculated in the frame of nonsuperfluid hydrodynamics,
is approximately 4 times larger than $\tau_{\rm b+s}$ 
determined self-consistently.
This difference arises because to plot the dotted curve
we used the formulas of Sec.\ \ref{Sec:damping_general}
in which $W_{\rm sfl}=V_{\rm sfl}=0$.  
As $T^\infty$ grows, however, 
the difference in two ways of calculating $\tau_{\rm b+s}$
rapidly decreases because the SFL-region becomes smaller 
and hence its contribution to $\tau_{\rm b+s}$
becomes less and less pronounced.
%

4. Unlike the radial and $p$-modes,
the agreement between dotted and solid lines 
for normal $f$-modes is very good
[see Fig.\ \ref{Fig:tau_nonrad}(b, c)],
which means that for these modes 
use of the nonsuperfluid hydrodynamics (far from resonances) 
is well justified.
The reason for such a good agreement of damping times
is related to a relatively weak compression-decompression
of matter in the course of the $f$-type oscillations. 
As a consequence, for the normal $f$-modes
the source $\delta \mu_{{\rm norm} \, l}$ in Eq.\ (\ref{sfl3})
is small, so that far from the resonances
$\delta \mu \approx 0$
and the superfluid degrees of freedom are almost not excited
[$W_{\rm sfl} \approx V_{\rm sfl} \approx 0$, 
see Eqs.\ (\ref{xisfl}), (\ref{xisfl2}), and (\ref{sfl2})].
This result confirms, extends and, we think, 
provides a deeper understanding, 
of the results previously obtained in a Newtonian framework 
by, e.g., \cite{lm94} and \cite{agh09}.

5. At $T^\infty \rightarrow T_{\rm cn \, max}^\infty=6 \times 10^8$~K 
one can observe the rapid increasing of $\tau_{\rm b+s}$ 
for superfluid modes in Fig.\ \ref{Fig:tau_rad}.
It is bounded from above by $\tau_{\rm bulk}$ and is related with
the tendency of $\tau_{\rm shear}$ to grow to infinity in this limit.
Such a behaviour of $\tau_{\rm shear}$
was discussed in detail in \cite{kg11}
and is specific for model 1 of nucleon superfluidity.

\section{Summary}
\label{Sec:summary}

In this paper we, for the first time, 
self-consistently analyze 
the effects of 
nucleon superfluidity on
damping of oscillations 
of nonrotating {\it general relativistic} NSs.
Our main results are summarized below.

1. The analytic formulas are derived 
for the oscillation energy $E_{\rm mech}$ (\ref{Emech2})
and for the characteristic damping times
$\tau_{\rm bulk}$ (\ref{taubulk2}) and $\tau_{\rm shear}$ (\ref{taushear2})
due to the bulk and shear viscosities.
These expressions are valid for oscillations of arbitrary multipolarity $l$.
The expression (\ref{Emech2}) for $E_{\rm mech}$
is the generalization of the formula (26) of \cite{tc67},
written for a nonsuperfluid NS.
The expressions (\ref{taubulk2}) and (\ref{taushear2})
are the generalizations, to the case of superfluidity,
of the formulas (5) and (6) in \cite{cls90}. 
Notice that the damping times, 
calculated using the formulas of \cite{cls90} 
appear to be 2 times smaller
than our $\tau_{\rm bulk}$ and $\tau_{\rm shear}$, 
calculated from Eqs.\ (\ref{taubulk2}) and (\ref{taushear2})
under assumption that superfluid degrees of freedom are suppressed
(i.e., $W_{\rm sfl}=V_{\rm sfl}=0$).

2. An approximate method is developed in detail and applied,
which allows one to easily determine the eigenfrequencies and eigenfunctions
of an oscillating superfluid NS, 
provided that they are known 
for a normal (nonsuperfluid) star of the same mass
(see Sec.\ \ref{SubSec:strategy}).
The method is based on the approximate decoupling of equations
describing superfluid and normal oscillation modes
and exploits the ideas first formulated in \cite{gk11, cg11}.

3. Using radial oscillations as an example, 
and adopting the simplified model 1 
of nucleon superfluidity (Fig.\ \ref{Fig:Tc1}), 
we demonstrate that this method leads to
oscillation frequencies and characteristic damping times 
that agree well with the results of exact calculation.

4. The approximate method of Sec.\ \ref{SubSec:strategy}
is applied to study nonradial oscillations of a superfluid NS
assuming the realistic model 2 of nucleon superfluidity 
(Fig.\ \ref{Fig:Tc2}).
A number of normal and superfluid oscillation modes 
with multipolarities $l=0,\ldots,3$ are considered.
In particular, the following normal modes are analyzed:
$F$-mode for $l=0$, $p_1$-mode for $l=1$, 
$f$- and $p_1$-modes for $l=2$ and $3$.

It is demonstrated that:
 
($i$) As a rule, for any given normal mode 
(whose frequency $\sigma$ coincides with the corresponding 
frequency of a nonsuperfluid NS and does not depend 
on the internal redshifted stellar temperature $T^\infty$)
the viscous damping time 
$\tau_{\rm b+s} \equiv (\tau_{\rm bulk}^{-1} +\tau_{\rm shear}^{-1})^{-1}$
is one order of magnitude {\it greater}
than $\tau_{\rm b+s}$ for those superfluid modes 
that can intersect the normal mode
in the $\sigma-T^\infty$ plane.
This effect is non-local 
(occurs only after integration over the NS volume)
and is determined by a number of factors
(see item 2 of Sec.\ \ref{SubSec:damping}).

($ii$) The function $\tau_{\rm b+s}(T^\infty)$ for any normal mode 
contains {\it resonance features}. 
In resonance points the frequency $\sigma$ of a normal mode
coincides with that
of some of the superfluid modes
(their $\sigma$ depend on $T^\infty$).
When passing a resonance (e.g., with growing $T^\infty$),
$\tau_{\rm b+s}$ initially rapidly decreases
(by 1--2 orders of magnitude)
until it reaches the value of $\tau_{\rm b+s}$ for this superfluid mode
and, after that, it increases again 
(see Figs.\ \ref{Fig:tau_rad}(d), 
\ref{Fig:tau_nonrad}, and \ref{Fig:tau_l2p1}).

($iii$) Resonance features (spikes) appear only 
in the approximate treatment of Sec.\ \ref{Sec:approach},
in which the normal and superfluid modes intersect 
at resonance points
[see, e.g., Fig.\ \ref{Fig:w_rad}(a)].
In the exact solution instead of crossings one has {\it avoided crossings}
of modes [Fig.\ \ref{Fig:w_rad}(b)].
Near avoided crossings any real mode changes its behaviour 
from normal-like to superfluid-like (and vice versa).
As a result, instead of spikes 
one has a very rapid step-like decreasing (increasing) of $\tau_{\rm b+s}$ 
[cf. Figs.\ \ref{Fig:tau_rad}(d) and \ref{Fig:tau_rad}(e)].
 
($iv$) Sufficiently far from the resonances 
$\tau_{\rm b+s}$ for normal radial and $p$-modes, 
determined self-consistently employing  
the hydrodynamics of a superfluid liquid, 
can differ several fold from $\tau_{\rm b+s}$, 
calculated using the ordinary normal-fluid hydrodynamics
(but accounting for the effects of superfluidity on 
the shear and bulk viscosities).
The latter approximation is often adopted in the literature
devoted to oscillations of NSs.

($v$) In contrast to radial and $p$-modes,
for $f$-modes far from the resonances,
use of the ordinary hydrodynamics of nonsuperfluid liquid 
for calculation of $\tau_{\rm b+s}$ is well justified.
The reason is that for $f$-type oscillations
the imbalance $\delta \mu$ of chemical potentials
is relatively small 
(matter does not compress significantly during oscillations).
Thus, superfluid degrees of freedom are almost not excited
(see Secs.\ \ref{SubSec:strategy} and \ref{SubSec:damping}).

($vi$) Since for $f$-modes far from the resonances $\delta \mu$ is small
(that is, deviation from the beta-equilibrium is weak),
bulk viscous damping of $f$-modes
is suppressed in comparison to $p$-modes.

Though here we only considered oscillations of superfluid nonrotating NSs, 
we expect that the main conclusions of this work
will also remain (mostly) unchanged for rotating NSs.
Our results indicate that 
dissipative evolution of oscillating NSs
may follow quite different scenarios 
than those usually considered in the literature.
This is especially true if one is interested in the combined
analysis of damping of oscillations and thermal evolution of a NS
or in the analysis of instability windows, that is 
the values of $T^\infty$ and rotation frequency at 
which a star becomes unstable with respect 
to the emission of gravitational waves 
(e.g., the $r$-mode instability, see \citealt*{andersson98, fm98}).
These issues are extremely interesting and important, 
but 
we left them beyond the scope of the present paper
and will address the related topics in our subsequent publication.

\section*{Acknowledgments}

This study was supported 
by the Dynasty Foundation, 
Ministry of Education and Science of Russian Federation 
(contract No. 11.G34.31.0001 
with SPbSPU and leading scientist G.G. Pavlov,
and agreement No. 8409, 2012), 
RFBR (11-02-00253-a, 12-02-31270-mol-a), 
FASI (grant NSh-4035.2012.2), 
RF president programme (grant MK-857.2012.2),
by the RAS presidium programme `Support for young scientists',
and by CompStar, a Research Networking Programme of the European Science Foundation.

\appendix
\section[]{Boundary conditions to equation (98)}
\label{appA}

Equation (\ref{sfl3}) should be solved in the region of a NS core
where neutrons are superfluid (SFL-region). 
If the NS centre is occupied by the neutron superfluidity, 
then for regularity of the solution at $r \rightarrow 0$
it is necessary that
\begin{equation}
\delta \mu_{l} \propto r^l.
\label{req0}
\end{equation}
The conditions at the boundary of the SFL-region
follow from the requirement of the absence 
of particle transfer (baryons and electrons)
through the interface.
One obtains from 
the definitions (\ref{X0})--(\ref{je2}) 
\begin{equation}
{\pmb X}_{\bot}=0,
\label{Xbot}
\end{equation}
where ${\pmb X}_{\bot}$ is the component
of the vector $X^j$ perpendicular to the interface. 
To rewrite Eq.\ (\ref{Xbot}) in terms of $\delta \mu_l(r)$, 
it is necessary to consider two possibilities:

($i$) The boundary (one of the boundaries)
between the SFL-region and nonsuperfluid matter 
lies inside the core and is defined by the condition 
$T=T_{\rm cn}(R_{\rm b})$ 
[$R_{\rm b}$ is the radial coordinate of the boundary].
Then at the boundary $Y_{\rm nn}(R_{\rm b})=Y_{\rm np}(R_{\rm b})=0$ 
and from Eqs.\ (\ref{sfl2}) and (\ref{sfl3}) one has
\begin{equation}
\delta \mu_l^\prime=\frac{{\rm e^{\lambda -\nu/2} \, \omega^2}}{h^\prime \, \mathfrak{B}}\,
(\delta \mu_l - \delta \mu_{{\rm norm} \, l}).
\label{gran1}
\end{equation}

($ii$) Outer boundary of the SFL-region coincides with the crust-core interface 
($R_{\rm b}=R_{\rm cc}$).
In that case $T < T_{\rm cn}(R_{\rm cc})$ 
[that is $Y_{\rm nn}(R_{\rm cc})$ and $Y_{\rm np}(R_{\rm cc})$ are non-zero]
and from Eq.\ (\ref{sfl2}) it follows that
\begin{equation}
\delta \mu_{l}^\prime(R_{\rm cc})=0.
\label{gran2}
\end{equation}
The conditions (\ref{req0})--(\ref{gran2})
are necessary and sufficient 
for solving Eq.\ (\ref{sfl3}).


\label{lastpage}

\end{document}